# Hypothetical bias in stated choice experiments: Part I. Integrative synthesis of empirical evidence and conceptualisation of external validity

Milad Haghani[1,*], Michiel C. J. Bliemer[1], John M. Rose[2], Harmen Oppewal[3], Emily Lancsar[4]

[1] Institute of Transport and Logistics Studies, The University of Sydney Business School, The University of Sydney, Australia
[2] Centre for Business Intelligence and Data Analytics, UTS Business School, University of Technology Sydney, Australia
[3] Department of Marketing, Monash Business School, Monash University, Australia
[4] Department of Health Services Research and Policy, Research School of Population Health, Australian National University, Australia

[*]corresponding author: milad.haghani@sydney.edu.au

**Abstract**

The notion of hypothetical bias (HB) constitutes, arguably, the most fundamental issue in relation to the use of hypothetical survey methods. Whether or to what extent choices of survey participants and subsequent inferred estimates translate to real-world settings continues to be debated. While HB has been extensively studied in the broader context of contingent valuation, it is much less understood in relation to choice experiments (CE). This paper reviews the empirical evidence for HB in CE in various fields of applied economics and presents an integrative framework for how HB relates to external validity. Results suggest mixed evidence on the prevalence, extent and direction of HB as well as considerable context and measurement dependency. While HB is found to be an undeniable issue when conducting CEs, the empirical evidence on HB does not render CEs unable to represent real-world preferences. While health-related choice experiments often find negligible degrees of HB, experiments in consumer behaviour and transport domains suggest that significant degrees of HB are ubiquitous. Assessments of bias in environmental valuation studies provide mixed evidence. Also, across these disciplines many studies display HB in their total willingness to pay estimates and opt-in rates but not in their hypothetical marginal rates of substitution (subject to scale correction). Further, recent findings in psychology and brain imaging studies suggest neurocognitive mechanisms underlying HB that may explain some of the discrepancies and unexpected findings in the mainstream CE literature. The review also observes how the variety of operational definitions of HB prohibits consistent measurement of HB in CE. The paper further identifies major sources and catalogues explanations of HB as well as possible moderating factors. Finally, it explains how HB represents one component of the wider concept of external validity.

## 1. Introduction

Do responses to hypothetical choice scenarios allow measuring preferences and predicting choices in real-world settings? This question—which is commonly referred to as the problem of *hypothetical bias* (HB)—arguably is the most fundamental question regarding the legitimacy of choice experiments (CEs) and the usefulness of CEs in policy making and cost-benefit analysis. A particular issue regarding CEs—also referred to as stated choice experiments (SCEs), discrete choice experiments (DCEs), or choice-based conjoint (CBC) in the literature—is the estimation of measures such as willingness to pay (WTP) or willingness to accept (WTA) and the extent to which the values obtained for these measures in hypothetical settings correspond to their values in real-world settings. In the absence of an inherent behavioural realism in hypothetical choice data, the use of CEs would be fruitless, regardless of the methodological advancements in capturing econometric phenomena using sophisticated model structures or the improvements in statistical efficiency of choice surveys. The question is not unique to choice modellers as it, in fact, represents the broader issue of generalisability which has been debated by social scientists, including experimental economists, for decades (Charness and Fehr, 2015; Herbst and Mas, 2015; Levitt and List, 2005, 2007).

The *validity* of a study concerns the extent to which the study's results represent and apply to the true state of some observed phenomenon. It is often discussed in terms of internal validity and external validity (Cook and Campbell, 1979). *Internal validity* refers to whether observed effects represent causal effects. CEs generally have high internal validity because the analyst can control for many aspects in the data collection including random allocation of participants to conditions. *External validity* (EV) refers to the generalisability of results beyond the study setting. As will be further defined later, HB is the bias in choice model estimates that results when data are collected in a hypothetical setting instead of in a more realistic setting. HB henceforth relates to EV and in fact is one aspect of EV.

CE's are typically administered as part of a questionnaire in a lab or in a survey. Lab settings allow maximising experimental control but as Levitt and List (2007) point out in their seminal paper, behaviour in the lab can be influenced not just by monetary incentives but by a range of other factors including social and ethical considerations (Kahneman and Knetsch, 1992). Surveys are low-cost and widely applicable instruments to study human preferences and decision-making and are often essential in policy making (Hainmueller et al., 2015). While there are many reasons why behaviour exhibited in a lab or survey may differ from real-world behaviour, their versatility in terms of representing choice settings under highly controlled conditions at low cost have justifiably prompted scholars to try to better understand the nuance surrounding the issue of generalisability beyond lab and survey settings and to identify ways to improve it (Falk and Heckman, 2009).

Gaining a nuanced understanding of EV of stated preference (SP) survey outcomes requires empirical investigations into the issue of HB. While ample empirical investigations exist with respect to *contingent valuation* (CV) surveys, auctions, or referenda (Loomis, 2014; Shogren, 2006; Vossler and Evans, 2009), research in relation to CEs has been much more limited (Ben-Akiva et al., 2019; Hensher, 2010; Lancsar and Swait, 2014; Quaife et al., 2018; Rakotonarivo et al., 2016). This is partly attributable to CE constituting a relatively young branch of SP methods, as most of the CE literature has accumulated only in more recent years and within a shorter time than, for example, CV (Ben-Akiva et al., 2019; Janssen et al., 2017; Lancsar and Louviere, 2008; Mandeville et al., 2014; Mühlbacher and Johnson, 2016). The extensive set of empirical evidence on HB in CV experiments has enabled economists to conduct meta-analytical studies, estimate the likely magnitude of HB and better understand what experimental factors and protocols aggravate HB (Carson et al., 1996; List and Gallet, 2001; Little and Berrens, 2004; Little et al., 2012; Murphy et al., 2005; Penn and Hu, 2019; Penn and Hu, 2018; Schmidt and Bijmolt, 2019). While these meta-analyses adopted different methods of calibration and different sets of variables, their findings suggest that HB can affect the estimates of WTP or WTA by a factor between 1.2 (Schmidt and Bijmolt, 2019) and 3.13 (Little and Berrens, 2004). A range of factors has been recognised to potentially affect the magnitude of the bias, including the type of good (public or private), the type of design (within- or between-subject), the elicitation method (e.g., choice versus open-ended CV) or the measurement type (WTP versus WTA ([Guzman and Kolstad, 2007](#))[1]). These meta-studies suggest that indirect elicitation methods, in particular CEs, may be less susceptible to HB than direct hypothetical valuation questions ([Murphy et al., 2005](#); [Penn and Hu, 2018](#)). However, like the influence of most of the abovementioned factors, the promising effect of the choice elicitation format on HB has not been unanimously supported by all meta data. [Little et al. (2012)](#) suggest that statistical evidence does not support the argument that CEs are less prone to HB than other typical CV approaches and [Schmidt and Bijmolt (2019)](#) find that indirect methods overestimate WTP more than direct methods.

The CE domain offers fewer empirical investigations of HB because real-world data counterparts of hypothetical choice data is often challenging or even impossible to obtain. A CE can have novel alternatives or attribute levels that do not (yet) exist in current markets, hence no revealed preference (RP) data may exist as a benchmark for evaluating HB. Studies that have reported such investigations appeared across different disciplines in which SP applications are common. In addition, different terminologies have been used to characterise such efforts. Many studies refer to EV ([Herriges et al., 1999](#); [Krucien et al., 2015](#); [Lancsar and Swait, 2014](#); [Louviere, 1988](#); [Louviere and Timmermans, 1992](#); [Quaife et al., 2018](#); [Rogers and Soopramanien, 2009](#)), while other directly refer to HB ([Buckell and Hess, 2019](#); [Hensher, 2010](#)). This has further hindered the acquisition of a holistic understanding of the extent to which HB occurs.

Furthermore, while HB in CV studies could be consistently operationalised—often by directly comparing the hypothetically stated WTP with actual payments in the form of a ratio—such a direct comparison is not possible for CEs and as such, the operationalisation of HB in CE studies has been variable, using a multitude of estimates and prediction measures that can be inferred from CEs. Depending on the main purpose of a CE, HB may relate to differences in marginal rates of substitution, opt-in rates, total WTP (TWTP), marginal WTP (MWTP), market shares or other measures. Also, cost is not always an attribute in CEs, essentially rendering the notion of WTP in terms of money (as a focal point for defining/measuring HB) meaningless. This lack of universality in characterising, measuring and reporting HB in CEs, as well as the lack of clarity about what benchmark best qualifies as a basis for evaluating HB (e.g., whether to use non-experimental revealed preference (RP) data or incentivised SP), in conjunction with the relatively small number of empirical investigations, have so far hindered a comprehensive analysis and synthesis of the evidence on this topic.

To address this lack of an integrative multidisciplinary investigation specific to CEs, we provide a comprehensive detailed review of the empirical findings on HB in CEs in each of the environmental, consumer, health, and transport economic domains based on systematised search strategy. We also investigate the variation in definitions of HB/EV across existing studies and propose a definition that is consistent across these domains and that can be operationalised. A list of the most relevant explanations and sources of HB in CEs is also presented, along with the moderating factors that could influence the magnitude of HB in CEs. We also visit the fields of psychology and cognitive neuroscience to seek insights regarding HB in CEs. We discuss how recently emerged evidence from experimental psychology and brain imaging studies can help delineate the HB as a phenomenon and help explain some of the unexpected or counterintuitive findings in the CE literature. Finally, Finally, recognising the variety of terminologies and dimensions of validity ([Khan, 2011](#); [Kimberlin and Winterstein, 2008](#); [McQuarrie, 2004](#)), we provide a unifying definition of HB for CEs and clarify the distinction between HB and EV by suggesting that HB may be best characterised as a component of the broader notion of EV.

The current paper presents Part I of our review of HB in CEs. In a separate paper presented as Part II ([Haghani et al., 2020](#)), we focus on the effectiveness of bias mitigation strategies. Part II will show how recognition of the sources of HB can be instrumental in determining best bias-mitigation strategies in a CE application.

---

[1]With respect to WTP and WTA, WTA values have been consistently found to be larger in magnitude than WTP outcome. Whilst economic theory attributes differences between WTP and WTA to possible income effects, HB has been suggested as a possible contributing cause (e.g., Guzman and Kolstad (2007)). With this caveat, we make no further distinction between the two and use WTP as a generic term to represent both.

The outline of this paper is as follows. In Section 2 we describe the search strategies and analysed datasets. Section 3 reviews empirical evidence on HB in CEs across various disciplines of applied economics. Section 4 discusses empirical evidence on HB from psychology and neuroimaging studies, while various sources of HB are discussed in Section 5. Section 6 presents a conceptualisation of EV and HB in CEs. Section 7 presents a summary and discussion and draws final conclusions.

## 2. Data and methods

Our search strategy was devised to comprehensively identify studies that have undertaken empirical investigations of HB in CEs. Two *general datasets* of references on HB were generated using the following search terms across the title, abstract and keywords in the Web of Science (WoS) Core Collection[2] (without restrictions on years of publications or document types):

General dataset A (429 publications):   "hypothetical bias"

General dataset B (235 publications):   ("stated choice*" OR "stated preference*" OR "choice experiment*" OR "discrete choice*" OR ("conjoint analysis" AND "choice*") OR "conjoint choice" OR "choice based conjoint") AND ("hypothetical bias" OR "external validity")

Publications in both datasets were examined individually and filtered against a set of inclusion and exclusion criteria to generate a *core dataset* consisting of 57 peer-reviewed journal articles reporting on empirical testings of HB in CEs. In order to qualify for inclusion, the article had to report an empirical comparison. Studies with no empirical data/investigation on HB (e.g., review papers or purely conceptual studies), or those unrelated to CEs (e.g., investigations of HB in CV) were excluded. The study had to have reported comparisons of discrete choice modelling measures of at least one kind (e.g., parameter estimates, willingness to pay, market share prediction) based on two datasets: one in hypothetical choice settings, and the other in an incentivised, or less hypothetical or real-world choice setting to qualify for the core dataset. For example, discrete choice studies that had made use of both stated and revealed choices were included only if they reported explicit comparisons or at least had separate but comparable model estimates across the two data types. Those that exclusively focused on combining stated and revealed choices as a way of data enrichment, without explicitly testing for HB, did not meet the inclusion criteria (e.g., Helveston et al. (2018)). A similar approach was taken with respect to studies of economic choice where incentive-compatible experimental methods were adopted but no comparison with a pure hypothetical treatment was reported (e.g., Yang et al. (2018), Gilmour et al. (2019)). Tests of EV or HB in setting other than CEs, such as CV (e.g., Bouma and Koetse (2019)) or auctions (e.g., Furno et al. (2019)) were also excluded. When the abstract did not contain sufficient information for making this determination, the full text was inspected.

Additional search strategies were adopted to ensure comprehensiveness of the core dataset. A series of less structured Google Scholar searches was undertaken, and a limited forward and backward expansion strategy was employed to enrich the search. The most recent articles in each dataset were singled out and their lists of references were individually checked in backward searching. Furthermore, a number of most relevant or most influential studies (based on citation counts) in each category were singled out and their citing articles were examined in forward searching.

Each item in the core dataset was individually reviewed, extracting the choice context, method of realism that had been set as the benchmark for HB evaluation, inclusion of an opt-out option in the design, type of respondents (e.g., students, households), basis of comparison (i.e., between-subject/split-sample, within-subject/repeated-sample design), model estimation method, comparison metrics, HB findings, and the main highlights of each study. In cases where the study identified significant bias based on certain metrics and negligible bias based on other relevant metrics, the finding of the study was deemed "mixed". Definitions of HB were also recorded for further analysis. Our search strategy also produced a *supplementary dataset* consisting of a set of less conventional studies related to HB in the domains of psychology and cognitive neuroscience. Items in this dataset were individually analysed to unmask possible implications for the mainstream CE literature.

## 3. Empirical evidence on hypothetical bias in choice experiments

In this section, we conduct a detailed analysis of the 58 peer-reviewed journal articles in the core dataset. These articles cover four disciplines in applied economics: environmental and resource economics (n=15, listed in Appendix A); consumer economics (n=18, Appendix B); health economics (n=10, Appendix C); and transport economics (n=14 , Appendix D).

### 3.1. Evidence from environmental and resource economics

Earliest investigations of HB in CEs have been reported in the environmental and resource economics literature, where SP methods in general and CEs in particular have traditionally been instrumental in determining valuations of non-market goods. In conducting such tests of bias, some studies (n=3) have been based on comparisons with equivalent self-reported RP surveys (Adamowicz et al., 1994; Adamowicz et al., 1997; Whitehead et al., 2008). Early studies in this category have shown that once differences in utility scales associated with SP and RP datasets are accounted for, inferences made based on SP and RP observations could yield similar outcomes (Adamowicz et al., 1994; Adamowicz et al., 1997).

Another cohort of studies (n=13) compared purely hypothetical discrete choice surveys with equivalent versions that included randomly drawn binding choices sets as tests of HB (Carlsson and Martinsson, 2001; Johansson-Stenman and Svedsäter, 2003). The underlying assumption is that in the latter condition, the monetary commitment associated with choices will make participants behave more truthful, making their responses unbiased or less biased. Similar to Adamowicz et al. (1997), a number of these studies failed to observe significant differences between non-incentivised and incentivised hypothetical choices in both within-subject (Carlsson and Martinsson, 2001) and between-subject testings (Cameron et al., 2002), although such evidence did not remain unchallenged as some subsequent testings produced evidence of HB (Ready et al., 2010) or mixed evidence (Whitehead et al., 2008). For example, in the context of beach access, through combining SP and RP data, Whitehead et al. (2008) were able to reliably estimate elasticity and consumer surplus using SP data (Brynjolfsson et al., 2019), while SP estimates of the number of trips were biased.

Other environmental studies have suggested the potential role of some moderating factors. Brown and Taylor (2000), for example, suggested that HB in valuation of a public good associated with male respondents, measured in the form of hypothetical WTP, was three times larger than that of females. Johansson-Stenman and Svedsäter (2008) also identified larger HB in between-subject designs than in equivalent within-subject designs. To explain this finding, they argued that people make an effort to appear consistent across their hypothetical and real (potentially binding) choices, a phenomenon often referred to as *cognitive consistency/dissonance* (McGuire, 1960) in psychology. In a follow-up study, Johansson-Stenman and Svedsäter (2012) investigated HB measured as the difference between MWTP in purely hypothetical and real-money treatments, across two experiments using a between-subjects design. One was in the context of a moral public good (contributions to a wildlife campaign) and the other was in a private morally neutral (amoral) choice context (a resultant voucher). While there was significant bias in the moral good context, no significant bias was observed in the morally neutral experiment. They theorised that, in addition to the benefits associated with the good itself, people derive utility from portraying a *positive self-image* which in part could explain why in SP settings, people tend to overstate their WTP for goods that exhibit an ethical component. They further argued that, given this theory, valuations inferred from real-money treatments involving moral goods may not be free of bias. In other words, experimental settings that put people under the assumption of contributing to a good cause may induce a positive bias, as they are likely to create a *warm glow* effect (Andreoni, 1990; Nunes and Schokkaert, 2003).

The study of Taylor et al. (2010) produced evidence that largely supports the theory of Johansson-Stenman and Svedsäter (2012) with respect to the connection between magnitude of HB and moral implications of the good. They observed no significant bias in MWTP estimates across hypothetical and binding choice treatments for a private good, while HB was significantly larger for their public good. They also explored the influence of two provision rules of binding choices, namely plurality vote and posted price market, on hypothetical choices for a public good (planting trees in public spaces) while comparing those with choices in a no provision rule condition. The notion of provision rule is also discussed by Vossler et al. (2012). They employed a game theoretic framework with field experiments designed to analyse incentive-compatibility properties of CEs, concluding that it is possible to elicit truthful preferences for non-traded public goods in hypothetical choice settings provided that participants perceive their decisions as having more than weak chance of influencing policy (i.e., *policy consequentiality*, Zawojska et al. (2019)).

As mentioned above, CEs in the context of environmental policy valuations often deal with non-tradable non-market public goods. As such, studies in this domain have been almost invariably limited to having participants pay the chosen amount of *donation* based on a randomly drawn choice set, as one viable option to induce incentive-compatibility. These payments are, in most cases, out of the monetary endowment participants receive in the lab, although exceptions also exist where people paid with their own money (List et al., 2006). Therefore, the *payment vehicle* is often, by default, assumed to be a direct donation. The study of Svenningsen and Jacobsen (2018) discusses the potential role of payment vehicle on HB in such laboratory settings. Their testing in the context of climate policies suggested negligible difference between a *hypothetical tax* and a *hypothetical donation*, as reflected in stated WTP estimates.

## 3.2. Evidence from consumer economics

Empirical testings of HB in CEs in the field of consumer economics (n=18) have been conducted predominantly in relation to food and beverage choice, with only few exceptions (Araña and León, 2013; Luchini and Watson, 2014; Wlömert and Eggers, 2016). Similar to the vast majority of environmental and resource valuation studies, the predominant method of measuring bias has been comparing purely hypothetical choices with financially consequential treatments where the individual would commit to purchasing the good that he/she has chosen in a randomly drawn choice set. Empirical evidence on the existence of HB in this domain is mixed though the majority of investigations (n=11) found strong evidence of significant bias.

In a pioneering empirical study of HB in consumer choice, Lusk and Schroeder (2004) observed that hypothetical choices of beef ribeye steak resulted in a significant overestimation of the probability of purchase, and thus of the total WTP, though no significant bias was associated with marginal WTP. The observation of a significantly overestimated opt-in rate (of purchase) is reminiscent of the work by Ready et al. (2010), who in a quasi-public good valuation (wildfire rehabilitation donation) experiment observed an opt-in rate three times larger in the hypothetical treatment than in the equivalent real-payment treatment. It is also in striking agreement with Alfnes et al. (2006), who in their experiment on salmon attributes observed overestimated hypothetical total WTP estimates, but no significant bias for marginal estimates.

In order to improve EV, it has been suggested that CEs should be *incentive-aligned* (Ding, 2007; Ding et al., 2005), also referred to as *incentive-compatible* (Lloyd-Smith and Adamowicz, 2018; Mørkbak et al., 2014). Ding et al. (2005) delineate the mechanism required according to the *induced value theory* of Smith (1976). This includes *monotonicity* (subjects must prefer more reward over less), *salience* (the reward amount must depend on actions of the subject) and *dominance* (utility influences other than that of the reward medium should be kept minimal). They also emphasise the role of salience of the experimental treatment. By comparing conventional and incentive-compatible CEs in the context of food consumption, they demonstrate the superiority of incentive-compatible methods to its conventional counterpart in predicting out-of-sample observations, hence adding to the evidence for the existence of HB in CEs. Wlömert and Eggers (2016) produced similar findings in the context of consumer adoption of a new music streaming service, showing that observations associated with an incentive-aligned CE better predicted real-market observations from the same individuals five month after service launch. Despite this evidence (Dong et al., 2010; Miller et al., 2011), it seems broadly agreed that incentive-alignment of CEs to remedy HB is not a universal solution and has the major limitation that its "application is limited to contexts where at least one concept of the research object can be rewarded after the experiment" (Wlömert and Eggers, 2016).

Consumer economic experiments have more often included an opt-out option in their experiments of HB than did environmental studies. This option was in some cases presented in the form of a "status quo" in otherwise forced choices (Sanjuán-López and Resano-Ezcaray, 2020). A closer look at the details of these experiments (see Appendix B) demonstrates that the inclusion of such an option does not guarantee bias-free estimates. In fact, in 8 out of the 11 experiments in this category that detected a significant HB, a status-quo or opt-out option had been included (Alemu and Olsen, 2018; Ding et al., 2005; Liebe et al., 2019; Luchini and Watson, 2014; Moser et al., 2013; Sanjuán-López and Resano-Ezcaray, 2020; Wlömert and Eggers, 2016; Wuepper et al., 2019).

Another technique to reduce HB is making the choice setting as tangible and relatable as possible for the respondents. This is often referred to as improving the *ecological validity* of experiments (Rossetti and Hurtubia, 2020). This has, for example, motivated some to test the effectiveness of virtual reality techniques (Matthews et al., 2017; Meißner et al., 2019; Romero et al., 2017; Rossetti and Hurtubia, 2020) and enhanced graphics and pictures (de Bekker-Grob et al., 2015; Haghani et al., 2015b; Ladenburg and Olsen, 2014; Rizzi et al., 2012; Tilley et al., 2016; Varela et al., 2014; Vass et al., 2019) in presenting choice alternatives as opposed to presenting attribute levels in words or numbers. Along this line, the study of Yue and Tong (2009) conducted field experiments on food product purchases with and without economic incentives, and tested if using real products instead of their pictures (in the hypothetical treatment) would make a difference in terms of reducing HB. Their observations confirm this hypothesis to a large degree.

*Information salience* is another factor that has been suggested as a possible explanation for HB in CEs (Haghani and Sarvi, 2017). This has also been tested in experiments on consumer food choices. Aoki et al. (2010) investigated how provision of information about food products (e.g., presence of additives) influences consumer choices, and how such impact would differ across purely hypothetical surveys and realistic experimental treatments. They observed that information provision on taste had a more pronounced effect in the real setting (where the subject had to purchase and consume the food), whereas health-risk-related information had a greater impact on choices in the hypothetical situation.

The existing literature offers another explanation that is less discussed with respect to HB. This concerns the fact that one-off hypothetical choice surveys consider only a cross-section of time and exclude the possibility of choice makers' learning and adjusting of choices (Sun and Morwitz, 2010) in dynamic adaptive real-world markets. This could be labelled as lack of *dynamic learning, adjustability and adaptation* and has close connection to the notion of *temporal stability* which has increasingly been mentioned in recent studies on CEs (Bliem et al., 2012; Doiron and Yoo, 2017; Lew and Wallmo, 2017; Liebe et al., 2016; Oppewal et al., 2010; Price et al., 2017; Rigby et al., 2016; Schaafsma et al., 2014). In the context of consumer preferences for corporate social responsibility, Araña and León (2013) investigated what they referred to as *dynamic hypothetical bias* by comparing predicted market shares based on a CE with real market shares. Their study showed that while predictions for initial periods were very accurate, as time evolved, real market shares displayed patterns that could not be reproduced by predictions based on hypothetical data.

As another study offering rather unfavourable evidence of the *reliability* of CEs, Luchini and Watson (2014) conducted an induced-value private-good CE and reported that it failed to elicit payoff-maximising choices and that neither increasing the salience of the attribute levels nor engaging subjects with monetary incentives could improve reliability. A related observation is the one by Meginnis et al. (2018) who found evidence of *strategic bias*, where respondents deliberately misrepresented their preferences to influence the decision-making process/policy in a study using induced-value CEs.

### 3.3. Evidence from health economics

The use of CEs has gained much traction and is now a well-established and accepted method among health economists to inform policy making (Clark et al., 2014; de Bekker-Grob et al., 2020; de Bekker-Grob et al., 2019; de Bekker-Grob et al., 2012; Ghijben et al., 2017; Lancsar and Louviere, 2008; Mandeville et al., 2014; Mühlbacher and Johnson, 2016; Payne and Elliott, 2005; Ryan and Gerard, 2003; Ryan et al., 2007). Compared to other domains of applied economics that we considered here, health economists have undertaken relatively few HB testings (n=10), likely due to the fact that the use of CEs in health is relatively new. The landscape of empirical evidence on HB in health economics also presents distinct differences to that of environmental and consumer economics. Firstly, while environmental and consumer economists replicate payment and policy consequentiality in their laboratory settings, empirical testings of HB in health economics are predominantly based on actual/real choices (n=9). These tests of bias/validity (Janssen et al., 2017; Lancsar and Swait, 2014) have resorted to a variety of real-world choices as benchmarks for evaluating the EV of CEs, including the actual uptake, choice or recommendation of a (new) medicine (Linley and Hughes, 2013), treatment (Krucien et al., 2015), vaccinations (Lambooij et al., 2015), diagnosis methods (Strauss et al., 2018), screening methods (Ryan and Watson, 2009), life-style interventions (Salampessy et al., 2015) and health-care plans (Kesternich et al., 2013). These efforts have concerned the choices of patients, the choices of physicians (Mark and Swait, 2004) or health authorities (Linley and Hughes, 2013). While HB in environmental and consumer studies has most often been evaluated based on marginal or total WTP, in health studies a variety of benchmarks have been reported. In addition to the usual WTP (for health-related goods and services) and parameter estimate measures (Mark and Swait, 2004), evaluations of EV based on sensitivity and specificity (Linley and Hughes, 2013), uptake rate or choice probability (Telser and Zweifel, 2007) and market shares (Buckell and Hess, 2019) have frequently been reported. The majority of these studies included an opt-out option (n=7). Health economic assessments of HB have also invariably been conducted using within-subject designs (n=10), whereas the environmental and consumer experiments present a mixture of within and between-subject assessments. Further, unlike the environmental and consumer economic studies, in health an accumulation of evidence points overwhelmingly (n=7) to the absence or to only a negligible effect of HB (Kesternich et al., 2013; Linley and Hughes, 2013; Mark and Swait, 2004; Mohammadi et al., 2017; Salampessy et al., 2015; Strauss et al., 2018; Telser and Zweifel, 2007).

A pioneering assessment of HB in health is presented by Mark and Swait (2004). Using a combination of SP and RP data on physicians' prescriptions of alcoholism medications, they suggest that parameters associated with the two data sources are equal, up to scale. Acknowledging various dimensions of EV as discussed in a broader context of social science studies, Telser and Zweifel (2007) differentiate between two variants of EV, i.e., *convergent validity* ("whether different methods that are designed to generate information about the same theoretical construct have convergent results"), and *criterion validity* ("whether the results of a method correspond with an external criterion", for example "comparing WTP calculated in a DCE with WTP observed in actual choices"). Recent reviews reveal that these terminologies, with certain variations in definitions, have gained acceptance in the health (Janssen et al., 2017) as well as the environmental domain (Hoyos, 2010; Rakotonarivo et al., 2016). Telser and Zweifel (2007) further highlighted the dimension of *reliability* as the attribute "that requires that measurements can be reproduced at least on average" (see the discussion on these terminologies and conceptualisation of EV in Section 6). Empirical testings reported by Telser and Zweifel (2007) in the context of hip protector choice for elderly provided strong evidence in favour of EV of CEs.

Consistent with assessments of HB in other fields, studies in health have also observed significant scale differences when comparing models of SP and RP data (Buckell and Hess, 2019; Kesternich et al., 2013). Among all differences across empirical evidence of HB in various fields, the scale factor seems to be a unifying piece of observation upon which there is relative consensus (Adamowicz et al., 1994; Adamowicz et al., 1997;

Alemu and Olsen, 2018; Haghani and Sarvi, 2017; Mark and Swait, 2004; Peer et al., 2014; Small et al., 2005; Wardman, 1991).

One aspect that is rather unique to the studies of HB in health as opposed to other domains is the focus on within-subject comparisons across an individual's hypothetical and actual health-related choices. In the case of within-subject comparisons, one can measure *sensitivity,* defined as "true positive rate" and an indicator of opt-in behaviour, and *septicity*, defined as "true negative rate" and an indicator of opt-out behaviour, see Quaife et al. (2018), as alternative model-free benchmarks for the assessment of bias (Linley and Hughes, 2013; Salampessy et al., 2015). Based on existing evidence, health-related CEs seem to have performed generally well in terms of these two measures of EV, in particular in terms of sensitivity (Quaife et al., 2018). In other words, the evidence suggests that the predictive value of CEs is higher for positive choice (opt-in) as opposed to negative choice (opt-out) (Lambooij et al., 2015).

### 3.4. Evidence from transport economics

There exist many applications of CE methods in transport (Hensher, 2010). Transport, along with health, are the two domains where SP methods have been most frequently used in demand estimation and attribute valuation of transport services (Basu and Hunt, 2012; Latinopoulos et al., 2017), infrastructure (Egbendewe-Mondzozo et al., 2010), facilities (Qin et al., 2017; Tilahun et al., 2007), and aspects of travel behaviour, human mobility and freight transport (Fowkes, 2007; Larranaga et al., 2017; Shams et al., 2017). Table 1 presents examples of transport studies using CEs.

**Table 1** Example transport studies using CEs (empirical investigations of HB in bold).

| Topic | Study |
| --- | --- |
| Acceptability and uptake of autonomous vehicles | Gkartzonikas and Gkritza (2019); Wang and Zhao (2019) |
| Driver behaviour | **Beck et al. (2016); Fifer et al. (2014)** |
| Responses to advanced travel time information | Abdel-Aty et al. (1997); Dell'Orco and Marinelli (2017); Engelson and Fosgerau (2020); Tseng et al. (2013); Wardman et al. (1997); Zhao et al. (2019) |
| Valuation of travel time savings | Correia et al. (2019); Devarasetty et al. (2012) |
| Valuation of travel time reliability & schedule delay | Alonso-González et al. (2020); Kou et al. (2017); Tseng and Verhoef (2008) |
| Responses to pricing policies | Holguín-Veras and Allen (2013); Lindsey (2011); **Nielsen (2004)** |
| Journey planning behaviour | Brazil and Caulfield (2013) |
| Commuter mode choice behaviour | Ahern and Tapley (2008); **Ghosh (2001)**; Jin et al. (2020); **Li et al. (2018)**; Small et al. (2005) |
| Commuter route choice behaviour | **Brownstone et al. (2003); Brownstone and Small (2005)**; Pahlavani and Delavar (2014); Shahhoseini et al. (2015) |
| Schedule preferences | Abegaz et al. (2017); Börjesson (2008); Herring et al. (2019); Hjorth et al. (2015); **Peer et al. (2014)**; Peer et al. (2016); Wen and Yeh (2017) |
| Residential location choice | Krueger et al. (2019) |
| Industrial location choice | Leitham et al. (2000) |
| Parking location choice | Chaniotakis and Pel (2015); Qin et al. (2017) |
| Airline choice | Grigolon et al. (2012) |
| Explicit time purchase | **Hultkrantz and Savsin (2017); Krčál et al. (2019)** |
| Preferences for alternative-fuel vehicles | **Brownstone et al. (2000)**; Cherchi (2017); Cirillo et al. (2017); Danielis et al. (2020); Shin et al. (2015); Ziegler (2012); |
| Preferences for shared mobility services | Correia and Viegas (2011); de Luca and Di Pace (2015); Ho et al. (2020); Wu et al. (2019); Yan et al. (2019) |
| Preferences for parcel delivery services | Herring et al. (2019) |
| Pedestrian route choice behaviour | Guo and Loo (2013); Haghani and Sarvi (2016); Haghani et al. (2016); **Haghani and Sarvi (2018); Haghani and Sarvi (2019)** |

Similar to other fields, SP applications in transport often consider novel non-existing products, services or technologies such as new metro lines (Fujii and Gärling, 2003), in which case robust testings of HB could not be expected. While SP and RP data is often combined to improve fit and prediction of the models when both datastores exist (Brownstone et al., 2000; Train and Wilson, 2008; Tseng et al., 2013; Zhang et al., 2018), most studies using joint SP-RP data do not explicitly test for HB. Therefore, empirical investigations of HB are confined to a limited set of applications (n=14) highlighted in bold in Table 1 and synthesised details of these studies are available in Appendix D. These empirical assessments have mostly based on the value of travel time saving (VTTS) (Ghosh, 2001; Krčál et al., 2019; Li et al., 2018) or value of travel time reliability (Brownstone and Small, 2005; Small et al., 2005) as the primary benchmark of assessment, while also comparative investigations based on individual-level simulated probabilities (Fifer et al., 2014; Haghani and Sarvi, 2018) and aggregate market shares (Beck et al., 2013; Haghani and Sarvi, 2019) have been reported. The RP benchmark for these comparisons have been sourced from a variety of methods, including non-experimental real-world choices (Brownstone et al., 2003), self-reported household travel surveys (Li et al., 2018; Wong et al., 2020), GPS-based location tracking (Beck et al., 2016; Fifer et al., 2014; Nielsen, 2004; van Essen et al., 2020), and laboratory experiments (Haghani and Sarvi, 2018; Hultkrantz and Savsin, 2017; Isacsson, 2007; Krčál et al., 2019; Swärdh, 2008). These assessments have been mostly based on between-subject comparisons and mostly without using an opt-out alternative. In contrast to empirical evidence from health economics, transport-related investigations have almost invariably found significant bias.

A pioneering testing of HB was reported by Brownstone et al. (2003) who investigated drivers' WTP for reductions in travel time during a congested morning peak period. Their RP observations (obtained from a congestion pricing project in the United States) are among the very few RP datasets used for such evaluations that are free of experimental interventions. They found that VTTS estimates using RP data were substantially larger than estimates based on SP data. This contradicts the vast body of evidence on HB that have shown that people overstate their WTP in hypothetical settings (Sanjuán-López and Resano-Ezcaray, 2020). However, the finding by Brownstone et al. (2003) has to a large degree been replicated by Nielsen (2004) based on a road pricing experiment in Denmark. Krčál et al. (2019) reported on laboratory experiments comparing equivalent SP and RP values associated with unexpected waiting times. In their design, SP and RP settings were completely identical with the only difference that RP choices were treated consequential in terms of waiting times and monetary incentives. They observed substantial degrees of downward HB, with the SP value being 70 percent of the RP estimate. Similar evidence of downward HB has been demonstrated by others ((Brownstone et al., 2003; Chiu and Guevara, 2019; Ghosh, 2001; Hultkrantz and Savsin, 2017; Isacsson, 2007; Krčál et al., 2019; Nielsen, 2004; Small et al., 2005).

Other studies have found an opposite effect. Li et al. (2018) reported that while SP estimates for private car VTTS in an Australian mode choice study are close to their RP counterparts, the SP estimate associated with public transport is significantly higher than in RP. Peer et al. (2014) made a similar conclusion with respect to the value of early and late schedule delay. Li et al. (2018) suggested that using an SP design pivoted around a real-market alternative could help reduce this gap. This hypothesis was tested in laboratory experiments by Hultkrantz and Savsin (2017) who observed that referencing to a real experience/alternative reduced the estimate of VTTS. They concluded that such as effect, in case of a negative HB, can in fact magnify the bias as opposed to mitigating it.

Fifer et al. (2014) tested the presence of HB in an SP vs GPS-field-experiment comparison of motorist behaviour in response to exposure-based charging. Using a within-subject design of the experiment, they found HB at the individual level. This study along with a follow-up analysis (Beck et al., 2016) showed that common bias mitigation practices (e.g., Cheap Talk, Certainty Scale Calibration) are effective in reducing the embodied bias.

Only two transport studies found no significant HB. Interestingly, neither of these studies embodied any monetary component or attribute. In the context of pedestrian evacuation route choice, Haghani and Sarvi (2018) found high consistency between SP and RP data and their modelling outcomes. Apart from scale, the comparison suggested complete SP-RP consistency in terms of sign and statistical significance of the estimates. The outcomes also suggested that attributes may be more salient when scenarios are presented as pictures, which may exaggerate their relative effect when compared to RP estimates of the same attribute. The SP and RP datasets of this study were subsequently compared against an independent RP dataset in terms of aggregate system-level measures when the choice model was applied to simulate a market/system (Haghani and Sarvi, 2019). SP-RP differences were found to be even smaller at the aggregate level.

## 4. Empirical evidence on hypothetical bias from psychology and neuroimaging

Increasingly, psychologists and particularly cognitive neuroscientists have also shown an interest in HB as a specific form of EV and this has resulted in a series of neuro-cognitive experiments. This increased interest coincides with development of neuro-economics as a field that aims to better understand the mechanisms underlying choice and decision-making through brain signals (Camerer et al., 2004; Goetz and James III, 2008). Some authors have referred to these efforts as "opening the black box behind the economic behaviour" (Jiang and Wu, 2016; Yu and Zhou, 2007). While the design of these experiments may in some cases not necessarily conform to the conventional design of SP surveys, their findings have provided valuable insight regarding human decision-making including decisions in SP survey settings. The choice contexts explored by these studies vary from delay and probability discounting in choice (i.e. the tendency to de-value/discount the outcome of a choice as a function of the temporal delay or probability of its occurrence) (Bickel et al., 2009; Lawyer et al., 2011), to time-independent consumer purchase of (desired) goods (Grebitus et al., 2013; Kang et al., 2011), choice of aversive goods (Kang and Camerer, 2013) and social moral choices (FeldmanHall et al., 2012a; FeldmanHall et al., 2012b; Vlaev, 2012). These studies have also adopted comparisons across purely hypothetical and some form of consequential treatments, namely real money gain based on a random choice draw (Bickel et al., 2009), real purchase of products (Kang et al., 2011), administering real (mild) harm to another person (FeldmanHall et al., 2012a) or actual consumption of food (Kang and Camerer, 2013) in order to study the phenomenon. Similar to the general body of empirical SP studies, the evidence of HB produced by this cohort of studies is rather mixed.

A pioneering contribution is the work of Bickel et al. (2009), which used functional magnetic brain imaging (fMRI) to collect brain signals in the context of intertemporal choices, i.e. valuation of commodities with respect to the delay until their receipt (as a choice attribute), using fictive and real money gain. Their within-subject design suggested no significant difference in discounting between hypothetical and real conditions. At the level of brain imaging, the analyses showed neither any significant differences in activation of certain brain regions across hypothetical and real treatments, which corroborates the observations at the level of manifested choices. This observation concurs with a body of experimental studies comparing the discounting of hypothetical and potentially real outcomes in various choice contexts – such as substance use and impulse-control choices (Lawyer et al., 2011), choices involving monetary rewards (Madden et al., 2003; Robertson and Rasmussen, 2018) as well as delay and probability discounting of food (i.e., choices between a smaller, sooner or certain vs. a larger, delayed or probabilistic number of bites of food) (Robertson and Rasmussen, 2018) – which generally observed hypothetical-real consistency. However, evidence to the contrary or mixed evidence exists. Experiments of Green and Lawyer (2014), for example, revealed steeper delay and probability discounting of potentially real versus hypothetical cigarettes among smokers, while discounting with direct monetary outcomes in the hypothetical and potentially real treatments was observed to be equal.

Neuroimaging experiments of consumer choice by Kang et al. (2011) also revealed evidence at the level of brain activity that, to some degree, explains and reflects on the existing mixture of findings on HB. Participants in an MRI machine were shown images of common consumer products (e.g., hand watches, backpacks etc) with fixed prices (informed by a pre-trial survey of the same participants) and were asked to indicate whether they would buy the product at the shown price. In the real choice treatments, subjects were instructed that one of the trials where the subject chose to "buy" the product will be chosen at random and that the subject is obliged to purchase the product out of their participation reward money. It was observed that decision times were substantially faster for real choices than for hypothetical choices when the *decision value* (i.e., the subject's hypothetical WTP for the item minus the posted price) was positive. For negative decision values, the difference in reaction times across hypothetical and real choices was insignificant. At the level of brain activity, the data suggested that hypothetical and real choices both activated common areas of the brain, but the strength of activation was higher in the real choice condition (Kang et al., 2011).

In another psychological investigation of HB in consumer choice, Grebitus et al. (2013) attempted to explain the phenomenon based on personality traits. They explored the role of personality in (real and hypothetical) choice and auction experiments and observed that personality traits play a more pronounced role in explaining HB in choice contexts than in auctions. The traits *agreeableness* and *neuroticism* were those with the least influence on choice-making behaviour while *agency* was the most important trait in a *big six* personality model of Lachman and Weaver (1997). Most personality traits had varying effects on WTP estimates across hypothetical and non-hypothetical experiments.

Neuroimaging studies of HB have also considered and explored the domain of moral choice. FeldmanHall et al. (2012a) investigated choices between financial self-benefit and preventing physical harm to others (in the form of an electric shock) in within-subject experimental settings under both hypothetical and real conditions while measuring decision-maker's fMRI brain signal. They observed that moral decisions with real consequences diverge from hypothetical moral choices, in that, people kept more money at the expense of inflicting pain to another person. They suggested this finding to be in stark contrast with the observations of Teper et al. (2011) showing that individuals in a moral action condition cheated significantly less on an actual math task than they had predicted they would (hence, acting more morally than they had expected). At the level of brain activity, while FeldmanHall et al. (2012a) identified a common neural network for both real and hypothetical conditions, they found distinct circuitry of activity specific to each condition. In a linked follow-up study however, FeldmanHall et al. (2012b) showed that by systematically enhancing the contextual information, it is possible to reduce the gap between responses in hypothetical and real situations. In doing so, they also emphasise the role of salience of information and showed that for example the influence of harm aversion diminishes as the impact of other motivational forces, such as significant financial gain, become more salient in the choice presentation. In light of this observation, FeldmanHall et al. (2012b) raise "questions about whether hypothetical moral decisions generated in response to decontextualized scenarios, act as a good proxy for real moral choices".

In another fMRI study Kang and Camerer (2013) revisited the well-established premise that hypothetical monetary valuations often overestimate real valuations by distinguishing between valuations of (desired/appetitive) "goods" and undesirable or (aversive) "bads", i.e., choices made to avoid aversive outcomes such as insurance purchase. They conducted experiments to see how much people are willing to pay to avoid eating an unpleasant food. They showed that the typical direction of HB may reverse in such cases as subjects seem to pay more to avoid bads when choices are real compared to when they make these evaluations on a purely hypothetical basis. At the level of brain activity, they observed that a real choice activates certain areas of the brain more strongly compared to a hypothetical situation. In justifying this finding, the psychological theory of "hot-cold empathy gap" (Loewenstein, 2005) has been invoked. This theorises a particular form of cognitive bias that makes people mis-predict/underestimate their own behaviour and preferences across affective states, meaning that "when people are in an affectively "cold" state, they fail to fully appreciate how "hot" states will affect their own preferences and behaviour" (Loewenstein, 2005). Therefore, the "atypical" observation of negative HB in valuations of travel time saving (VTTS) in transport economic studies (Brownstone et al., 2000; Brownstone et al., 2003; Brownstone and Small, 2005; Chiu and Guevara, 2019; Ghosh, 2001; Hultkrantz and Savsin, 2017; Isacsson, 2007; Krčál et al., 2019; Nielsen, 2004; Small et al., 2005) may not be so counter-intuitive after all, when contrasted with the typically positive/upward HB in valuations of private and public goods (Sanjuán-López and Resano-Ezcaray, 2020). In a context of commuting route choice for example, a commuter's willingness to pay toll could be regarded as the money paid to avoid an aversive bad (i.e., being stuck in heavy traffic or being late to the destination), an affective state that may be grossly discounted in the choice-maker's mind when making the choice in a purely hypothetical setting (i.e. when in the "cold" state). Whereas, when facing the actual possibility of facing heavy traffic versus avoiding such experience by paying more (i.e. when in a "hot" state), the individual may in fact be willing to be pay a larger amount to avoid the aversive experience.

Further details of the 12 psychological studies that investigated HB can be found in Appendix E.

## 5. Sources and explanations of hypothetical bias

### 5.1. Sources of hypothetical bias

Our review demonstrates that the justifications and explanations offered for the existence of HB are diverse and several theories have been proposed from different angles. Lee and Hwang (2016) state that "unfortunately, there is not yet a consensus on the underlying drivers of this hypothetical bias".

Many economists have suggested that HB arises simply due to the lack of *consequentiality* of hypothetical responses, or lack of *incentive compatibility* (Buckell et al., 2020; Lewis et al., 2018; Mørkbak et al., 2014). This itself embodies two different dimensions. First, unless certain measures are taken to make the hypothetical survey incentive-aligned (Ding et al., 2005), the respondents are no worse or better off than they were prior to the survey regardless of their stated responses to valuation scenarios, i.e. lack of *payment (individual) consequentiality* (Zawojska et al., 2019). Hence, they may not have a genuine reason to be truthful and price-sensitive, take their *budget constraints* into consideration (Ding et al., 2005) or put considerable *cognitive effort* into their responses (Wlömert and Eggers, 2016). Secondly, respondents may believe that their response has no real policy consequences and only serves the intellectual curiosity of some researchers, i.e. lack of *policy (societal) consequentiality*.

It has been suggested that the absence of payment consequentiality may result in *deceitful* answers. There are conceivable reasons and scenarios where a respondent may think that revealing their true preferences leads to unfavourable outcomes and therefore may mislead the investigator if they face zero liability for their stated responses (Frank et al., 2017; Thanos et al., 2011). For example, if respondents believe that a company is conducting the CE to determine an acceptable price for a new product, then they may strategically choose to appear more price-sensitive through their choices and deflate their WTP for the product. Or similarly, in a transport context, respondents may strategically select hypothetical route alternatives with no or low toll costs in opposition to new toll roads. This *deliberate misrepresentation* of behaviour, which is sometimes referred to as *strategic behaviour* (Lu et al., 2008; Meginnis et al., 2018), may similarly occur in CEs about public goods. A respondent may, for example, try to manipulate an agency into providing the good or service in question, and *free-ride* (Lusk et al., 2007; Throsby and Withers, 1986; Veisten and Navrud, 2006) on the actual donations of others (if the provision mechanism is voluntary donations). While it has been discussed that the structure of multi-attribute multi-alternative CEs may inherently make it more difficult for participants for adopt strategic behaviour compared to CV surveys, it appears that CEs could still be vulnerable to this issue (Burton, 2010; Thanos et al., 2011). This warrants the question whether an investigator should mask the true purpose of the survey when there exists the likelihood of *strategic response* (Lloyd-Smith and Adamowicz, 2018; Meginnis et al., 2018) or *protest response* (Ami et al., 2011; Meyerhoff and Liebe, 2008). Alternatively, one could inform respondents of the large sample size to which they are contributing in order to disincentivise strategic answers.

Another important reason suggested for the existence of HB is that, when the contexts of choice presents moral or prosocial elements, respondents may choose options that they believe makes them appear more socially desirable or altruistic to investigator(s). This could particularly be of concern when a provision of a public good with positive social implications (e.g. environmental conservation, improved public health, subsidised public transport) is the focus of the survey (Auger et al., 2003; Auger et al., 2007; Auger et al., 2008). Such *warm glow* effect (Andreoni, 1990; Nunes and Schokkaert, 2003) may make respondents overstate their WTP through their choices, particularly when payment consequentiality is lacking (Johansson-Stenman and Svedsäter, 2012). This *social desirability* effect is commonly described in the literature of survey applications (Champ and Welsh, 2006; Ding et al., 2005; Hainmueller et al., 2015; Leggett et al., 2003; Menapace and Raffaelli, 2020; Olynk et al., 2010; Sanjuán-López and Resano-Ezcaray, 2020; Smith et al., 2017; Svenningsen and Jacobsen, 2018).

Even when surveys are perceived consequential, and even if we assume that possible incentives for being strategic or socially desirable can be neutralised through effective measures, there will still be undeniable factors that could drive biased responses to an SP survey. The most prominent factor could be what psychologists have described as the *hot-cold empathy gap* phenomenon (Kang and Camerer, 2013; Loewenstein et al., 1998), a cognitive bias (or failure of perfect imagination) that inherently limits humans to correctly predict our future behaviour (Frederick et al., 2002; Loewenstein et al., 2003; Loewenstein and Schkade, 1999). This could be an important factor when we ask about preferences for goods and services that are non-existent or those with which the subject has no prior experience (e.g., self-driving vehicles). As discussed in Section 4, brain scanning studies have suggested that our brains do not respond the same way when we make decisions that seem hypothetical or far from the "here-and-now". From this perspective, contemporary versions of hot-cold empathy gaps in psychology tailored to CE could potentially become a new generation of HB testing. The convention in HB evaluation has so far been to compare a purely hypothetical setting with a counterpart setting that includes a form of payment.

Tailoring the choices to real payments, however, is not a universally applicable option for testing HB. Many CEs consider alternatives where financially binding stated choices are not practically feasible (e.g., car ownership choice, infrastructure investment choice). In cases where investigators have reason to believe that bias is mostly attributable to lack of experience or familiarity with the alternatives (Kealy et al., 1990; Lusk and Norwood, 2009; Zhao et al., 2011), lack of contextual tangibility (Yue and Tong, 2009) or even lack of dynamical learning and adaptation (Araña and León, 2013; Lusk and Norwood, 2009), one may consider testing subjects in a (relatively) "hot state" and compare outcomes with that of a pure hypothetical. This could be a space where technological tools such as driving simulators (Fayyaz et al., 2020; Hess et al., 2020) and virtual reality (Matthews et al., 2017; Meißner et al., 2019) could assist to produce more evidence on HB. For instance, since the inception of research on autonomous vehicles, an abundance of CEs have been conducted mainly focussing on estimating WTP (Gkartzonikas and Gkritza, 2019). It is not clear how the issue of HB in this context could possibly be tested using conventional approaches. However, one can envisage administering a survey to participants right after they had a test drive in an autonomous vehicle (i.e. in a "hot state") and comparing that with a control group, as a proxy test for HB. Further, certain CEs do not contain a cost attribute, for example when considering pedestrian behaviour (Haghani and Sarvi, 2017; Haghani et al., 2015a) or driver's reaction to variable message signs (Wardman et al., 1997; Zhao et al., 2019). Such CEs would understandably not entail any monetary trade-off, therefore, a criterion such as payment consequentiality as a way of testing HB would not be relevant. Instead, a comparable counterpart experimental design in a driving simulator or even in the field (if practical) could be a reasonable way forward in testing HB in such contexts.

Another issue that could potentially explain some experimental findings of HB in CEs is the role of *cognitive dissonance* (Alfnes et al., 2010; Izuma et al., 2010; Shultz et al., 1999) as another psychological phenomenon, particularly when using a within-subject design. The theory of cognitive dissonance is based on the notion that materialised actions/choices of people could affect their preferences. A typical cognitive dissonance experiment asks participants to rate a number of goods, make a number of choices between pairs of those goods, and then rate them again. It has been shown that, once a subject makes a difficult choice between two items that he/she prefers almost equally, his/her subsequent rating for the chosen (rejected) item increases (decreases) compared to the original rating. The theory postulates that holding two or more contradictory cognitions at the same time poses a psychological discomfort that people try to avoid and may even change preferences in order to do so. This tendency of humans to modify preferences to align them with past actions, i.e., the existence of choice-induced preference change, has also been demonstrated by neuroimaging studies (Sharot et al., 2009). This may play a confounding role when one tests HB in two subsequent experimental settings where participants' choices in the first experiment could potentially influence their preferences expressed in the second experiment. A similar argument is the one regarding 'coherent arbitrariness' as proposed by Ariely et al. (2003). A first choice, even when made on a random basis, will influence following choices with the result that preferences such as expressed in CEs appear as more stable (see also Schmidt and Bijmolt (2019)). Relatedly, it has been argued that choosers seek to justify their choices and so will try to appear consistent (Simonson, 1989).

The trade-off between attributes of alternatives often requires a considerable amount of cognitive effort, and may result in simplifying strategies such as choosing the opt-out option frequently (without considering the presented trade-off) (Wlömert and Eggers, 2016), *yea-saying* (Blamey et al., 1999; Blarney and Bennett, 2001; Holmes and Kramer, 1995) or *anchoring* (Mørkbak et al., 2010; Van Soest and Hurd, 2008) to minimise *cognitive effort* and *attention*. This highlights the role of *design artefacts* (Ladenburg, 2013) that have the potential to create or magnify bias. This may embody issues such as information salience (Aoki et al., 2010), attribute salience (Bogomolova et al., 2020; FeldmanHall et al., 2012b; Haghani and Sarvi, 2017, 2019), and survey mode or setting (Menegaki et al., 2016). A summary of the sources for HB is provided in Figure 1.

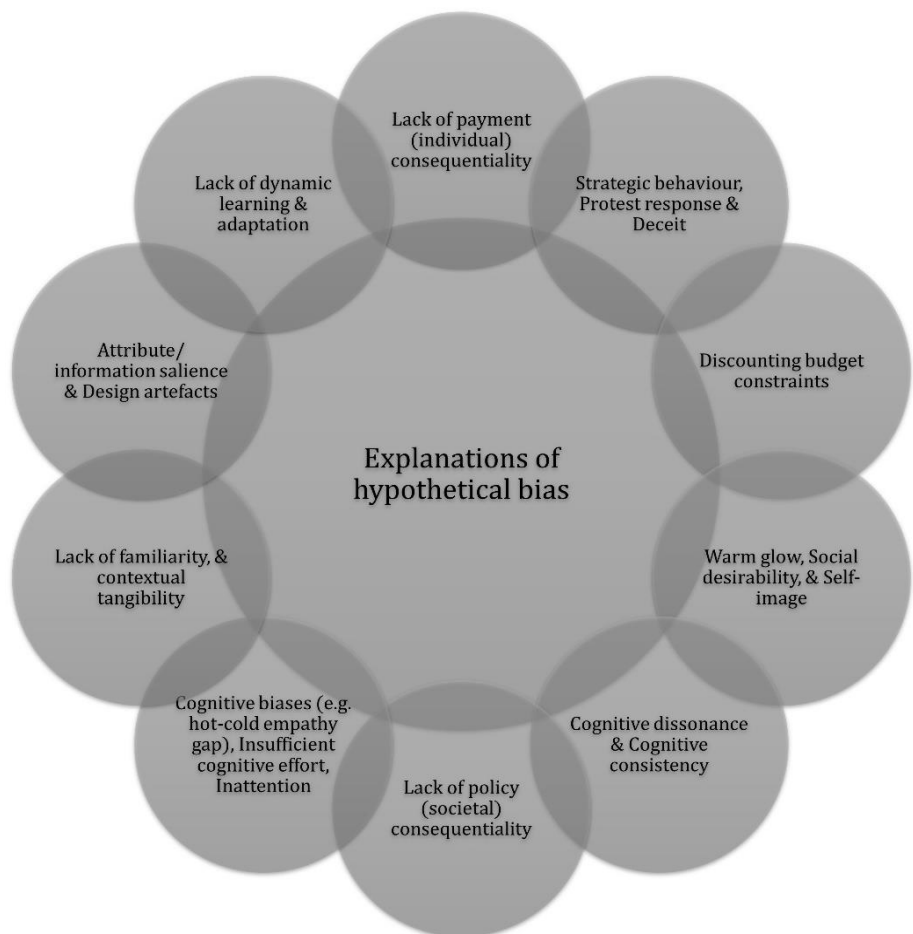

**Figure 1** Possible sources of HB in CE.

## 5.2. Moderating factors of hypothetical bias

The current literature has recognised a range of factors that may correlate with the extent of HB in CEs. We differentiate moderating factors from sources of bias because the effect of sources could be countered through proper measures, whereas moderating factors are inherent to the design of a survey and can only be corrected or accounted for by understanding their role in influencing the magnitude or direction of bias. Unlike the literature on CV, evidence on these moderating factors in CEs is limited and scattered, calling for more research into their effects.

*Individual characteristics* of participants in choice surveys have been recognised as a potential factor that could influence the magnitude of HB (Wuepper et al., 2019). This includes observable characteristics such as *gender* (Brown and Taylor, 2000; Johansson-Stenman and Svedsäter, 2012), as well as unobservable characteristics that should rather be enquired from participants through suitable supplementary questions, factors such as their *knowledge* and *familiarity* with the goods in question (Sanjuán-López and Resano-Ezcaray, 2020), or *personality traits* (Grebitus et al., 2013). Knowing, for example, what gender or what personality types are more susceptible to HB could itself lead to bias correction factors or methods. The *characteristics of the good/service* constitute another potential factor that may influence HB, with suggestions that *public* and *private* goods (Svenningsen and Jacobsen, 2018), goods/services with and without *social/moral components* (Johansson-Stenman and Svedsäter, 2012) or *desirable* versus *undesirable* goods (Aoki et al., 2010) may be differently prone to HB. Another factor that has been pointed out include the *stake size* and the range of the cost attribute (Isley et al., 2016; Ohler et al., 2000), suggesting that the larger the (hypothetical) monetary stake, the greater the HB. In public good valuation, the *payment vehicle* (e.g., tax versus donation) has also been suggested as a potential influencing factor (Penn and Hu, 2019; Svenningsen and Jacobsen, 2018).

## 6. Defining hypothetical bias in choice experiments and its relation with external validity

### 6.1. Conceptualisation and definition of hypothetical bias

In this section we propose a broad definition of HB in CEs that reflects, to the best possible extent, the overall view in the literature and the diversity of CE applications. Carson et al. (2014) points out that the concept of HB "has been defined in a number of inconsistent ways in the literature". Table 2 provides definitions from individual studies, which reveals that many HB definitions are very narrow by referring to specific measures, benchmark data source for comparison, and even bias direction, e.g. Brown and Taylor (2000) define HB as "overstating of true values for a public good when the payment decisions are not binding". In this section, we propose a unified and inclusive definition of HB for CEs that can be applied across disciplines and applications. Given that bias directions vary across disciplines and applications, we define HB as a *deviation* without a presupposition of the direction of the measured difference.

With respect to benchmark data source, definitions in Table 2 refer to "revealed", "real", "true" or "actual" to indicate choice observations in the real world. While this seems straightforward, such choices can only be observed in a *naturalistic* setting where behaviour of agents is observed without influencing their behaviour. For example, Robin et al. (2009) use field RP observations for the estimation of a pedestrian route choice model. Self-reported revealed choices may not reflect true behaviour due to potential warm glow and social desirability effects, protest responses, or limitations in memory to recall past choices. In some studies, participants are given devices to automatically record choices to avoid these issues, but the knowledge of being monitored may influence their behaviour. There are only very few studies that have used naturalistic observations as a benchmark for determining HB in CEs. Rather than using naturalistic observations as a benchmark, most studies use choice observations collected in a *more realistic setting* compared to the hypothetical setting. Experimental economists have long recognised the important role of "creating a lab in the field" (Viceisza, 2016) as a suitable compromise between controllability in the lab and realism of naturally-occurring data (Gneezy and Imas, 2017; List, 2007; Simester, 2017). From this perspective, tests of HB do not have to occur in naturalistic settings as in many cases true preferences may never be obtainable (Veisten and Navrud, 2006). Clearly, the variation in the degree of realism when testing for HB should be accounted for as a measure for the rigour of testing. However, even a minimal comparison across degrees of realism may give the analyst an idea of the possible direction and magnitude of HB.

Based on the literature, we identify five degrees of realism as shown in Figure 2. Class I data is collected in the least realistic setting and class V data is collected in the most realistic setting. Classes I to III refer to CEs where only classes I and II refer to hypothetical or laboratory settings, which may be prone to HB. Benchmark data sources range from classes II to V. In order to explain these classes, consider a route choice context with the aim to estimate VTTS. Class I data is collected in typical online stated choice surveys, while an example of class II data is provided in Fayyaz et al. (2020) who observed route choices in a driving simulation laboratory where participants experience travel times and are faced with real toll costs. Class III data could be obtained by instructing participants in the CE to make specific trips at specific times of day using their own car in the real world. Examples of Class IV data are self-reported route choices drawn on a map or map-matching GPS routes through devices installed in the cars of participants. Class V data could be collected by analysing mobile phone data obtained directly from mobile service providers, see e.g. Bwambale et al. (2019). The arrows indicate studies of HB across a hypothetical setting and a more realistic setting, where the line widths are proportional to the number of studies that have made this comparison (a dashed line means that no studies yet exist). All evaluations of HB to date have considered class I data in the hypothetical setting. Most evaluations of HB (n=39) consider class II data as the benchmark, mostly in the psychology, environmental economics, and consumer economics literature. Class III data is less frequently used as benchmark (n=7), mostly in the transport literature, while class IV data is used in 16 studies as benchmark, mostly in the health literature. Only 7 studies (in transport and consumer economics) have considered class V naturalistic data as benchmark. Benchmark data classes for each study can be found in Appendices A–E.

With respect to measures to determine HB, the vast majority of existing HB definitions pivot around economic terms such as "*payment*", "*market*", and most frequently, "*WTP*", with a focus on the disparity in monetary valuation of attributes. One could measure bias based on disaggregate metrics (such as sensitivity and specificity, individual total or marginal WTP or simulated probabilities for individual choices) as well as aggregate measures (such as scale-adjusted pairwise comparisons of parameter estimates, marginal rates of substitution, total or marginal WTP, elasticities, market shares). At a higher level of aggregation, one would also consider cases where discrete choice models are estimated to be incorporated as part of a multi-layer model (such as a traffic simulation model) to predict indirect ultra-aggregate measures (such as total travel time spent in the network). In other words, a unified definition of HB should not be restricted to a single measure. Given that different data sets have different measurement errors (or error variance), where class I data typically has the smallest measurement error and class V data the largest, one may need to account for scale differences in parameter estimates. Using WTP (if a cost or price attribute exists) or marginal rates of substitution (if no cost/price attribute exists) as a measure avoids this issue.

Based on the above analysis, we propose the following unified and inclusive definition of HB in CEs:

> *Hypothetical bias is the deviation in a predefined aggregate or disaggregate measure due to choice data being collected in a hypothetical setting instead of a more realistic (but not necessarily naturalistic) setting.*

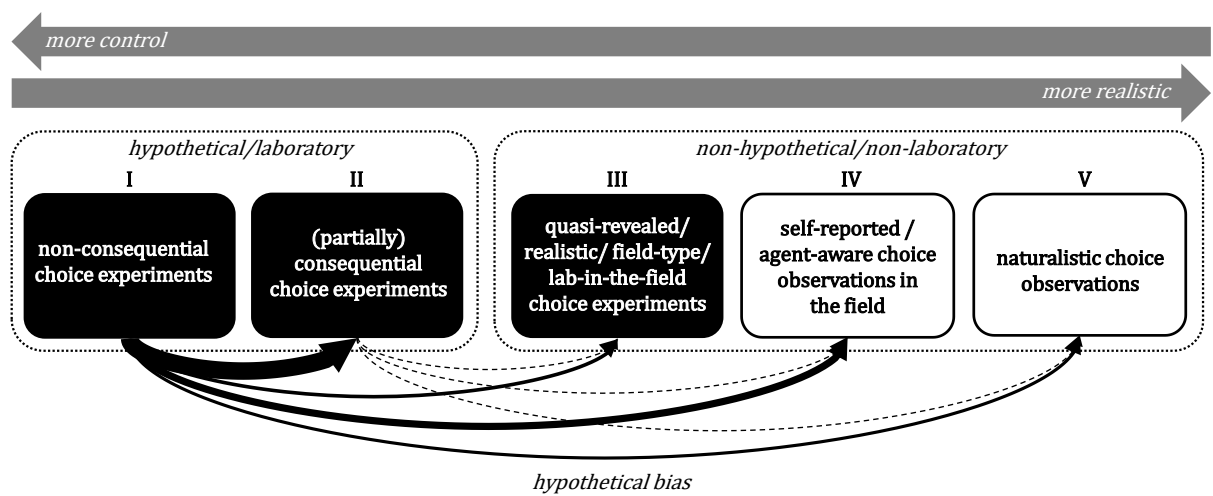

**Figure 2** Conceptualisation of HB across various degrees of realism. The thickness of continuous curved arrows is proportional to the number of empirical investigations in the CE literature that have used that method. Dotted curved arrows represent types of investigating HB that, thus far, have not been reported in the literature.

**Table 2** Definitions of hypothetical bias according to the existing literature.

| Reference | Definition |
|---|---|
| Aadland and Caplan (2003) | One definition of *hypothetical bias* encompasses any deviation of an individual's stated WTP that is due to the hypothetical nature of the good. |
| Asensio and Delmas (2015) | This distance between what people say they would do and what they actually do is referred to as *hypothetical bias*. |
| Börger and Hattam (2017) | There may still be a discrepancy, however, between whether individuals would pay in a real life situation and the amount that they would actually be willing to pay. Such *hypothetical bias*… |
| Brown and Taylor (2000) | Differences in responses under hypothetical and real conditions are attributed to *hypothetical bias*, which refers to the overstating of true values for a public good when the payment decisions are not binding. |
| Carlsson et al. (2010) | We will denote the difference between the real and hypothetical treatments as *hypothetical bias*. |
| Chavez et al. (2020) | Inconsistency between stated and revealed preferences is exacerbated in the absence of economic incentives (*hypothetical bias*). |
| Fifer et al. (2014) | The different choices made by individuals in hypothetical settings as opposed to those made in real life situations is often described as resulting from *hypothetical bias*. |
| Gracia (2014) | Most of these studies using RCEs have focused on studying *hypothetical bias* by comparing results from both the hypothetical and the non-hypothetical versions of choice experiment. |
| Hensher (2010) | The extent to which individuals might behave inconsistently, when they do not have to back up their choices with real commitments, is linked to the notion of *hypothetical bias*. |
| Kang and Camerer (2013) | *Hypothetical bias* is the common finding that hypothetical monetary values for "goods" are higher than real values. |
| Krčál et al. (2019) | Most studies find that the value [of travel time saving] elicited from SP data is substantially lower than the value elicited from RP studies. This gap is usually referred to as *hypothetical bias*. |
| Ku and Wu (2018) | *Hypothetical bias* is broadly known as the divergence between the hypothetical WTP and real values. |
| Lee and Hwang (2016) | One lingering concern over its [CVM] use, supported by a wealth of empirical evidence, is that respondents tend to overstate their WTP in hypothetical settings [omitted references]. Unfortunately, there is not yet a consensus on the underlying drivers of this "*hypothetical bias*". |
| Lewis et al. (2018) | Despite the wide use of choice experiments, they may be prone to *hypothetical bias*, as respondents do not have to support their choices with real commitments. |
| Li et al. (2018) | The difference between the results estimated from SP and RP data is commonly referred to as a source of *hypothetical bias*. |
| List and Gallet (2001) | we refer to "*hypothetical bias*" as the difference between hypothetical and actual statements of value, where actual statements of value are obtained from experiments with real economic commitments. |
| Little and Berrens (2004) | *Hypothetical bias* can be defined as the disparity between hypothetical statements and real values (or what an individual might actually pay for the provision of the good) |
| de-Magistris and Pascucci (2014) | We refer to hypothetical bias when individuals overstate their WTP in hypothetical settings, and then behave inconsistently when they do not have to back up their choice through a formation of real commitments |
| Mamkhezri et al. (2020) | One concern with the use of these methods is that stated preference surveys may be subject to bias, in particular *hypothetical bias* – the gap between WTP response to a hypothetical question and actual payment behaviour to a real incentive. |
| Moser et al. (2013) | SPs have often been found to differ from real preferences [omitted reference]. For example, estimated hypothetical willingness to pay (WTP) is generally higher than actual WTP for real goods, thus, providing evidence for the existence of '*hypothetical bias*'. |
| Murphy et al. (2005) | The hypothetical nature of these surveys – in both the payment for and provision of the good in question – can result in responses that are significantly greater than actual payments. This difference between stated and revealed values is often referred to as *hypothetical bias*. |
| Ozdemir (2015) | The difference between what people say they are willing to pay in a survey, and what people actually would pay using their own money in a laboratory or field experiment, is known as *hypothetical bias*. |
| Ready et al. (2010) | The difference between hypothetical values and actual payment values is referred to as *hypothetical bias*. |
| Rose et al. (2015) | …'*hypothetical bias*', a condition whereby respondents answering SC survey tasks respond in a manner other than how they would if faced with similar choices in real markets |
| Schmidt and Bijmolt (2019) | The difference between hypothetical willingness to pay and real willingness to pay is the *hypothetical bias*. |
| Strauss et al. (2018) | The main objection to SP data is *hypothetical bias* [omitted reference]. That is, what people say they will do and what people actually do can be very different. |
| Svenningsen and Jacobsen (2018) | The most prominent factor identified in this debate is the discussion regarding *hypothetical bias*, a term used to capture the idea that when faced with no real consequence of their choices, people will tend to overstate their willingness to pay for a certain good. |
| Tilley et al. (2016) | *Hypothetical bias* represents the potential divergence between real and hypothetical payments. |
| Vlaev (2012) | The only reliable difference between hypothetical and real responses in this research field is what is known as the *hypothetical bias*: people overstate hypothetical valuations. |
| Vossler et al. (2012) | Accumulated evidence, largely from laboratory experiments, shows systematic deviations between stated and revealed preferences [omitted references]. This is generally referred to as "*hypothetical bias*". |
| Wuepper et al. (2019) | *Hypothetical bias* can be defined as the difference between stated, hypothetical behaviour and actual behaviour in the real market |
| Yue and Tong (2009) | … hypothetical bias (the difference between what people say they will pay and what they would actually pay)… |

*6.2. External validity and its relation to hypothetical bias*

In the previous section, we demonstrated how the definition of HB shows structural variations across different existing studies and hence, the need for a unifying overarching definition that could embody HB in a broader variety of choice experimental contexts. We also observed, referring to the literature of CE, another layer of confusion resulting from HB often being presented as synonymous with EV. This begs the question whether the two are distinguishable at all and how the relationship between the two notions can be best characterised, within the specific context of CE. Table 3 lists several definitional quotes of EV derived from the literature of CE and, in conjunction with Table 2, shows how EV and HB have often been used interchangeably (Hensher, 2010; Lewis et al., 2018; Lusk and Schroeder, 2004; Strauss et al., 2018; Vossler et al., 2012).

In order to explain the difference between HB and EV, we need to revisit the broader concept of *validity* (Karren and Barringer, 2002; Toubia et al., 2003; Vossler et al., 2012) as a notion that far predates the use of SP methods and one that is commonly used in various areas of social and medical sciences to characterise qualities of *experiments* (Khan, 2011; Lynch Jr, 1982; Winer, 1999), and *measurement instruments* (Kimberlin and Winterstein, 2008)[3]. Upon inspection of the topic, one notes that there exists no consensus even in the terminologies that have been used to describe validity of behavioural experiments. This controversy is perhaps best captured by Campbell (1986) stating that "Confusion about the meaning of validity in quasi-experimental research can be addressed by carefully relabelling types of validity. Internal validity can more aptly be termed 'local molar causal validity.' More tentatively, the 'principle of proximal similarity' can be substituted for the concept of external validity." (p. 67). As a reflection the extent of debate surrounding these terminologies, in the same article entitled "Relabelling Internal and External Validity for Applied Social Scientists" Campbell (1986) even expresses dissatisfaction with the new terms proposed by his own article.

Setting aside the lack of existence of unanimity amongst social scientists in defining validity, a rather well-accepted (though not-universally-accepted) point of view that has also been adopted by studies in the CE literature is that the validity of an experiment can be considered from two different angles of *internal* and *external* validity (Janssen et al., 2017; Zawojska and Czajkowski, 2017). In general terms, internal validity concerns the ability of an experiment to reveal causal relations, a goal which oftentimes requires abstraction and simplification as well as experimental control (Cook and Campbell, 1979; McQuarrie, 2004). Internally valid experiments produce outcomes that are robust and replicable. In a specific CE context, internal validity is thought to mainly concern the key assumptions and quality of execution of a choice survey, the robustness of the design and the analytical rigour of data collection (Janssen et al., 2017; Johnson et al., 2019; Telser and Zweifel, 2007). A CE is deemed to score high on internal validity when it possesses two general qualities: (i) it fulfills the conditions of best design recommendations (Lancsar and Louviere, 2008; Lancsar and Swait, 2014; Louviere and Lancsar, 2009) and (ii) the responses of participants represent the theoretical assumptions of consumer demand theory (Johnson et al., 2011; Tervonen et al., 2018). Characteristics (i) and (ii) are typically labelled as *content* and *construct* (or *theoretical*) validity and represent two aspects of the broader notion of internal validity in CEs. While tests and criteria for assessing both aspects have been proposed in the literature (Janssen et al., 2017; Ryan et al., 2009), it is important to note that the literature of CE does not treat these two with equal significance, as recent studies have given more weight to the importance of content validity while recognising that construct validity in CE may not be of vital importance. In fact, it has been suggested that what we refer to as content validity has important bearings on the EV of CEs as their internal and external validity are fundamentally inseparable (Lancsar and Swait, 2014). Some authors have even suggested that, in certain circumstances, efforts to improve construct validity of a CE (e.g., by removing the so-called "irrational choices" (Miguel et al., 2005; Ryan et al., 2009) could come at the expense of the EV of outcomes (Lancsar and Louviere, 2006; Tervonen et al., 2018). Some authors have even regarded construct validity as a dimension independent from internal validity (Calder et al., 1982; McQuarrie, 2004). The overall view is that while we have clear recommendations of best practice for CE design and improving content validity (and thereby, EV), the ways to assess construct validity are not as clear. As aptly pointed out by Zawojska and Czajkowski (2017) this "requires well-defined theory as a reference point for the comparison of theoretical predictions and stated values" (p. 377) and that, as behavioural economics has shown us, "even in real markets, consumers are observed not necessarily to behave in line with this theory" (p. 377).

---

[3] While CEs can arguably be also regarded as instruments for indirectly measuring (inferring) preferences, here, we give more weight to the points of view and definitions that have originated from (behavioural) economic or psychological experiments, as opposed to those that concern the notion of validity in relation to measurement instruments, more common in clinical psychology and medical fields.

The discussion on the trade-off between external and internal validity has a long history in experimental behavioural research (Cook and Campbell, 1979) including consumer research (Calder et al., 1982; Lynch Jr, 1982) and economics (Chytilova and Maialeh, 2015). As pointed out by Schram (2005), the decision as to which is more important goes back to the primary purpose of the experiment. It has been suggested that internal validity could be the primary concern when designing an experiment for theory testing, whereas for those seeking empirical regularities, external validity could be a more important criterion (Schram, 2005). The views on this issue are, however, not unanimous and alternative views have been presented too (see, for example, Calder and Tybout (1999) or Calder et al. (1981). In the case of a CE, a relevant theoretical construct could be the marginal rate of substitution and the issue would be whether this construct as measured in the CE represents the same mechanism or trade-off as the one operating in real-world settings. So, what makes a CE externally valid? A view embodying different perspectives and characterisations of EV in CE would indicate that a CE is deemed externally valid when the following holds. (i) Different established methods measuring the same construct (e.g., different design versions of a CE, ranking of alternatives, hedonic pricing, or CV) produce similar (convergent) results; and (ii) its outcomes are fairly consistent with those of an external criterion assumed to be a good proxy for true preferences (e.g., those assumed to underlie an RP dataset, or observations from an equivalent incentive compatible or field experiment). These two conditions can be respectively referred to as *convergent* and *criterion* (predictive) validity. Similar to the two main dimensions of internal validity, whether convergent or criterion validity is of greater importance for assessing EV of a CE depends on the research goals. For example, given the ongoing debate as to whether CE or CV methods are more suitable for preference elicitation, one cannot necessarily interpret any disparity between WTP inferred from a CE with that obtained from a CV as an indication that the CE is less valid. Such lack of convergence, while seemingly only of theoretical interest, will make empirical results equivocal and hence may make them less trustworthy. Mostly however, when referring to HB, the researcher's interest is in criterion validity. In the case of CEs this dimension is inextricably related to what psychologists refer to as *ecological validity* (i.e. aspects dealing with the *environment* of an experiment, its conditions, settings and treatment) (Bracht and Glass, 1968; Rossetti and Hurtubia, 2020; Schmuckler, 2001).

The abovementioned characterisation recognises that while the most important aspect of the EV of a CE is HB, HB represents only one component of the wider concept of EV. A wider question is then also where within such a conceptualisation the issues of generalisability across population (i.e. population validity) (Bracht and Glass, 1968) and generalisation across time (i.e. temporal validity) (Munger, 2019) may fit. It seems they do not fit either category perfectly. Population validity (i.e. generalisation from an experimentally accessible population to a target population) is often deemed a component of EV but it is also a generic aspect of any experimental investigation involving human subjects and is not specific to CEs. One could, therefore, regard this a separate dimension of EV, in addition to convergent and criterion, when it comes to CEs. As to the generalisability across time, the issue is often labelled as *reliability* (Cook et al., 2007) or *temporal reliability* (Schaafsma et al., 2014) or *temporal stability* (Lew and Wallmo, 2017) and is generally treated independently from evaluations of validity (Rakotonarivo et al., 2016).

**Table 3** Definitions of (external) validity according to the existing literature.

| Reference | Definition |
|---|---|
| Araña and León (2013) | A potential problem of these [SP] methods is the likelihood of producing different results than those obtained with an actual market setting, thereby questioning their *validity* for assessing preferences and values. |
| Chang et al. (2009) | Although a great deal has been learned, there are very few studies examining the *external validity* of these methods. As such, scepticism surrounding stated and experimental willingness-to-pay values abounds |
| Hensher (2010) | Non-experiment *external validity* tests involving observation of choice activity in a natural environment, where the individuals do not know they are in an experiment, are rare. In contrast the majority of tests are a test of *external validity* between hypothetical and actual experiments. |
| Johansson-Stenman and Svedsäter (2003) | The ultimate *validity* test of SP methods is often considered to be the extent to which statements of maximum Willingness To Pay (WTP) correspond to real or actual payments…such *external validity* tests are rare. |
| Johansson-Stenman and Svedsäter (2008) | An obvious *validity* test of SP methods is to compare hypothetical statements with people's real willingness-to-pay (WTP). |
| Johansson-Stenman and Svedsäter (2012) | The extent to which WTP statements correspond with real-money payments is often seen as the ultimate *validity* test of SP methods. |
| Lewis et al. (2018) | Few choice experiments are able to validate their findings through *external validation* with a real market due to the difficulty in identifying a market valuing the same attributes. … Many authors have suggested that the reliability and *validity* of choice experiments should be tested through comparisons with real or simulated markets. |
| List et al. (2006) | While these other institutions have generally not performed well, in that hypothetical and actual behaviour has not perfectly matched, this study represents a first attempt in the field to provide a firm understanding of the *external validity* properties of the CE approach |
| Strauss et al. (2018) | … what people say they will do and what people actually do can be very different. This can, in some cases, lead to specious forecasts lacking *external validity*. |
| Telser and Zweifel (2007) | *External validity* refers to the generalizability of the results to other populations, settings and circumstances… In this paper we focus on two variants of *external validity*, viz. convergent and criterion validity. |
| Quaife et al. (2016) | There is a great need for future research on the *external validity* of DCEs, particularly empirical studies assessing predicted and revealed preferences of a representative sample of participants. |
| Vossler et al. (2012) | Accumulated evidence, largely from laboratory experiments, shows systematic deviations between stated and revealed preferences [omitted references]. … This lack of correspondence has raised questions about the criterion (i.e., *external*) *validity* of stated preferences. |

## 7. Summary and conclusions

As a commonly used method of economic valuation, CEs are widely employed across multiple disciplines. CE methods are essential tools for inferring human preferences in a large variety of applications, including monetary valuation of public environmental policies, price setting and demand estimation of novel consumer goods, estimation of patients' and physicians' various treatments and health care plans, and valuation of travel time savings and reliability. These constitute crucial inputs for cost-benefit analysis of policies and public projects. When researchers resort to CEs these are, more often than not, the only practical option for obtaining these estimates. Given the hypothetical nature CEs, a fundamental question is whether the estimates properly reflect preferences and WTP as they operate in more realistic or natural settings.

Given the prevalence of CEs in economic valuations and the increasing popularity of these methods, the issue of HB warrants more recognition and investigation. However, there are several factors that have so far hindered obtaining a holistic and nuanced understanding of this problem. First, empirical testings of HB are often difficult or even impossible to conduct in many choice contexts due to the lack of a benchmark that reflects "true" preferences. Secondly, the literature of HB in CEs is intertwined with that of a broader CV literature and this has made it hard to disentangle the evidence from CV methods and form an evidence-based knowledge of the issue of HB specific to CEs. Thirdly, empirical observations of HB in CEs are scattered across multiple fields where the contexts of choices or methods of evaluating HB could vastly differ. These fundamental differences result in different conclusions about the prevalence, magnitude and direction of the bias if only certain subsets of empirical investigations or certain domains are considered. We therefore conducted an integrative and multidisciplinary analysis of the HB in CE literature. As visually summarised in Figure 3, by referring solely to empirical studies of HB in the health domain, one may reach the conclusion that HB is perhaps not a big issue in CE, whereas the evidence from transport economics points to the contrary. However, a holistic analysis would suggest that for every study that has suggested HB to be negligible, there are nearly two studies that provide evidence to the opposite, making the issue of HB in CEs an undeniable issue.

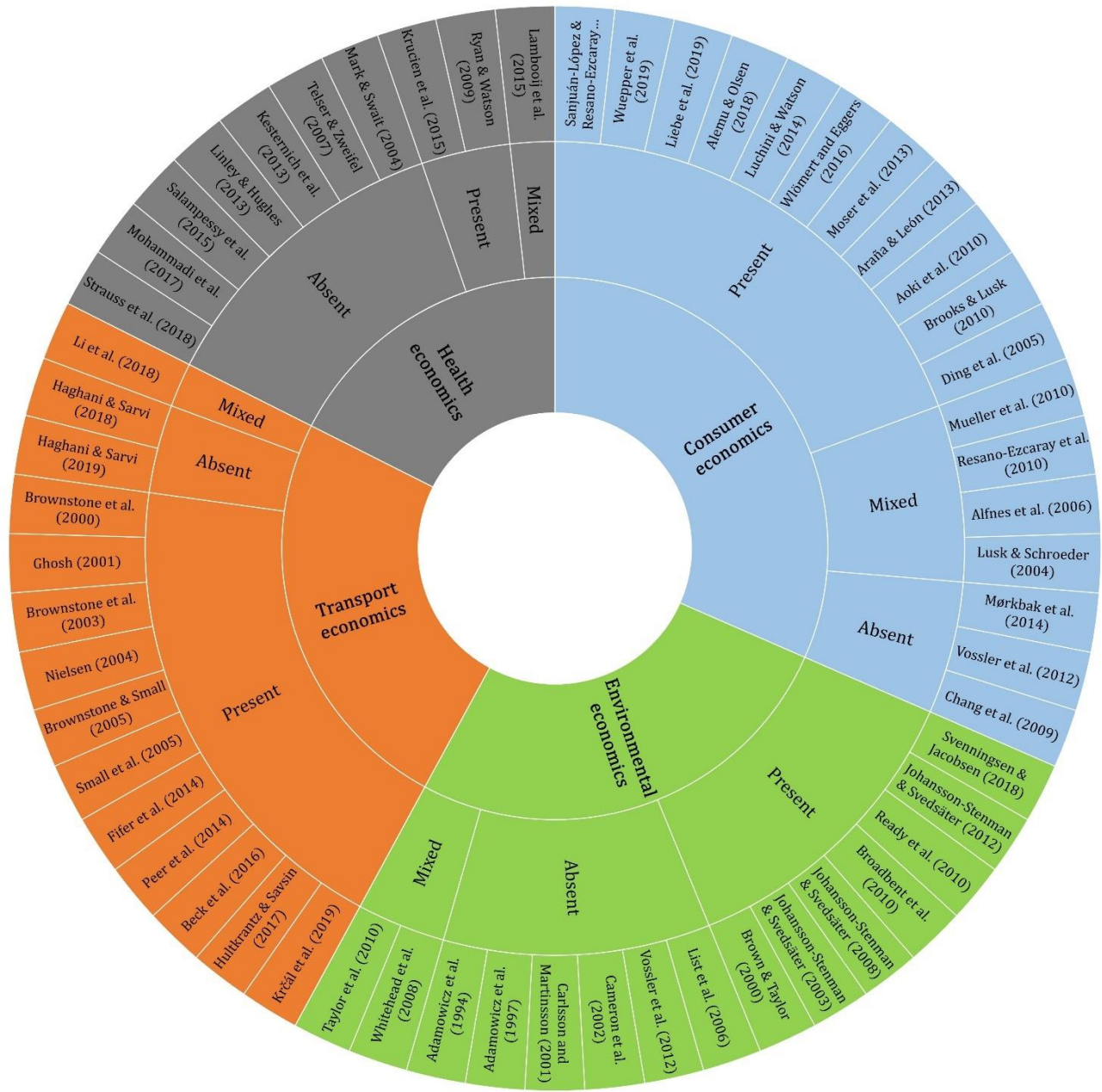

**Figure 3** The empirical studies of HB in CE from environmental, consumer, health and transport economic disciplines along with their main finding on HB, i.e. whether HB was "absent" or "present" or that the evidence was "mixed" (e.g., HB was significant based on one measure and negligible based on another).

Why is it that, unlike two other domains, most investigations in health contexts have found no evidence of significant HB? A reasonable speculation concerns the role of choice context and respondent involvement in moderating HB in CE. Conceivably, survey questions on health are perceived as more important, prompting patients/doctors to take experiments in health more seriously, than survey questions answered by consumers about, for instance, pasta. This means it is important to study HB in CEs in a context-specific and nuanced manner while considering potential moderating factors, instead of only asking the generic question whether HB exists in CEs. For example, while the evidence that has been produced in the transport domain points predominantly to the presence of significant HB (mostly in the form of a downward bias in monetary valuation of travel time savings), the two transport-related investigations where the choice did not entail any monetary trade-off (Haghani and Sarvi, 2018, 2019) did not find significant bias. Whether these observations indicate that hypothetical choices with no financial trade-off/implication or those with important personal consequences will be less likely to suffer from HB is a question that warrants more empirical investigation.

In order to reduce HB in CEs, it is important that one recognises the sources of HB and their likely prevalence in various CE applications such that mitigation strategies can be adopted. New insights from the behavioural sciences (including psychology and cognitive neuroscience) may be instrumental in developing a better understanding of the processes underlying HB and/or in explaining seemingly contradictory evidence in the literature. One particular reason that may have hindered testing of HB in many choice contexts—particularly in transport studies where the number of investigations of HB seems disproportional to the number of CEs—is the aim to use "true" preferences as the benchmark. We argued that, in the vast majority of applications, such a golden standard is simply not available or feasible. A pragmatic solution to this issue is testing across various degrees of realism (as shown in Figure 2). In fact, any CE known to be less susceptible to HB (e.g., in terms of payment consequentiality, contextual tangibility, etc) could be set as a benchmark for the evaluation of HB in a standard purely hypothetical version of the CE. For example, administering a choice survey aimed at estimating WTP for driverless cars right after the person has experienced the product, or actual trade-off of people's time with earned money in the lab while simulating their commute, could be as legitimate for testing HB as the random drawing of a financially binding choice set in a consumer choice application, which is an established and well-accepted method. Clearly, the more realistic the benchmark, the more rigorous the HB test will be. Such differential factors can be accounted for in possible subsequent meta-analyses of HB by assigning different weights to various testing methods. However, the absence of a perfect criterion, for example in the form of RP data, need not be deemed prohibitive for testing HB in choice contexts where this issue has remained underexplored.

## Acknowledgments

This research was funded by Australian Research Council grants DP150103299 and DP180103718.

## Appendix A: Summary of relevant studies in environmental economics

The table below presents a summary of studies in the discipline of environmental economics on hypothetical bias in choice experiments.

| Reference | Choice context | Benchmark data class | Opt-out | Sample | Within/ between sample | Estimation method | Measures of validity/bias | Sig. bias? | Conclusions/comments |
|---|---|---|---|---|---|---|---|---|---|
| Adamowicz et al. (1994) | Recreational site (water resources) | IV | Yes | Participants of water-based recreation | Within | -Multinomial logit -Scaled logit | -Parameter estimates -Predicted probabilities | No | Joint estimation that accounts for scale differences gives evidence that the underlying preferences inferred from both methods are similar. |
| Adamowicz et al. (1997) | Recreational site (moose hunting) | IV | Yes | Participants of hunting trips | Within | -Multinomial logit -Scaled logit | Parameter estimates | No | - The hypothesis of scale parameter equality between the SP and RPp (the perception version) could not be rejected - The hypothesis of scale parameter equality between the SP and RPo (the objective version) was rejected - The hypothesis of parameter equality between the SP and RP model could not be rejected |
| Broadbent et al. (2010) | Environmental donation | II | Yes | Students | Between | -Multinomial logit -Multinomial probit | MWTP | Yes | - CEs, like CV, also demonstrate HB - Internal validity tests indicated stability and transitivity |
| Brown and Taylor (2000) | Environmental donation | II | N.A. | Students | Between | Regression | -Parameter estimates -WTP | Yes | - Examines gender as a possible explanation of hypothetical bias - Gender differences exist in hypothetical valuation exercises, but not in real valuation exercises - Hypothetical bias is almost three times larger for males than for females |
| Carlsson and Martinsson (2001) | Environmental donation | II | No | Students | Within | -Multinomial logit -Scaled logit | -Parameter estimates -MWTP -Direct choice response comparisons | No | The hypothesis that preferences are the same in both experiments cannot be rejected |
| Cameron et al. (2002) | Environmental donation | II | Yes | Randomly selected households | Between | Scaled logit | WTP | No | - Simultaneous comparison of choices elicited by seven alternative preference elicitation methods, one RP and six SP - When different data types were pooled, it was not possible to reject the hypothesis of identical underlying preference for at least four elicitation methods |
| Johansson-Stenman and Svedsäter (2003) | Environmental donation | II | No | Students | Between | -Multinomial logit -Random effects logit -Fixed effects logit | MWTP | Yes | MWTP for contributions is highest when choices are hypothetical, lower for a follow-up real-money experiment, and lowest when real-money choices are made directly. |
| Johansson-Stenman and Svedsäter (2008) | Environmental donation | II | No | Students | Between & Within | Multinomial logit | MWTP | Yes | - Larger hypothetical bias in between subject than in a within-subject design<br>- People strive for consistency between their attitudes and behaviours |
| Johansson-Stenman and Svedsäter (2012) | Moral & neutral goods | II | No | Students | Between | - Multinomial logit<br>- Random effects logit<br>- Fixed effects logit | MWTP | Yes | - Develop and test a theoretical model to explain variations of hypothetical bias in the literature<br>- The model proposes that people derive utility from a positive self-image<br>- The model predicts that in SP people overstate their MWTP for moral goods, but not for morally neutral goods.<br>- No significant gender differences in hypothetical bias |
| List et al. (2006) | Public good & private good | II | Yes | Random sample of households + Participants of an sportscard show | Between | - Linear regression | Proportions | No | - Hypothetical CEs combined with "cheap talk" can yield credible estimates of the purchase decision<br>- No evidence of hypothetical bias when estimating marginal attribute values<br>- Responses across the real and hypothetical with cheap talk were not statistically different |
| Ready et al. (2010) | Quasi-public good (Wildfire rehabilitation donation) | II | Yes | Students | Between | Multinomial logit | - Opt-in rate<br>- WTP | Yes | - WTP for a public good estimated from SP choices was three times as large as WTP estimated from choices requiring actual payment<br>- The opt-in rate for SP choices was about three times as large as for choices involving real payments.<br>- The bias was related to the stated level of certainty of respondents |
| Svenningsen and Jacobsen (2018) | Moral good (climate policies) | II | Yes | Random sample from general population | Between & Within | Random parameter logit | - Parameter estimates<br>- MWTP | Yes | - Highlighting the significance of design components (e.g. payment vehicle, bid range) on the extent of the hypothetical bias |
| Taylor et al. (2010) | Public & private goods | II | Yes | Students | Between | Multinomial probit | - MWTP<br>- TWTP | Mixed | - For private good, MWTP from the hypothetical treatment is larger, but not statistically different than that of the binding choice treatment<br>- For public good, MWTP from the hypothetical treatment is much larger, and statistically different than that of the binding choice treatment |
| Vossler et al. (2012) | Tree planting project | II | No | Volunteer university employees | Between | Scaled logit | - Parameter estimates<br>- MWTP | No | - A large difference in the results from both hypothetical treatments compared to the real money treatment<br>- MWTP for donations was higher when subjects stated their own preferences but lower when they stated what they believe are others' preferences |
| Whitehead et al. (2008) | Recreation site | IV | No | Random sample of beachgoers | Within | Random effects Poisson | - Parameter estimates<br>- Elasticity<br>- Consumer surplus | Mixed | - Hypothetical bias affects the estimates of number of trips and slope coefficient<br>- Hypothetical bias does not affect elasticity or consumer surplus estimates |

*Sig. bias: Significant bias*
*HB: Hypothetical bias*
*WTP: Willingness to Pay*
*MWTP: Marginal willingness to pay*
*RP: Revealed Preference*

## Appendix B: Summary of relevant studies in consumer economics

The table below presents a summary of studies in the discipline of consumer economics and marketing on hypothetical bias in choice experiments.

| Reference | Choice context | Benchmark data class | Opt-out | Sample | Within/ between sample | Estimation method | Measures of validity/bias | Sig bias? | Conclusions/comments |
|---|---|---|---|---|---|---|---|---|---|
| Alemu and Olsen (2018) | Novel food products | II | Yes | Random sample of consumers | Between | Random parameter logit | -Parameter estimates -WTP -Market shares | Yes | - Larger utility scales for the hypothetical treatment - Full consistency in terms of the sign and significance of estimates - Higher marginal willingness-to-pay values in hypothetical than in non-hypothetical settings |
| Alfnes et al. (2006) | Food quality (Salmon fillets) | II | Yes | Random sample of consumers | Within | Mixed logit | -Parameter estimates -TWTP -MWTP | Mixed | The hypothetical choice scenarios overestimate the TWTP, but that the MWTP is not significantly different |
| Aoki et al. (2010) | Food additives | II | No | Students & commuters | Between | N.A. | WTP | Yes | - WTP values were lower in real experiment compared to the hypothetical survey - Information on flavour had more influence on choice behaviour in the real situation - Information on health risk had more influence on choice behaviour in the hypothetical situation |
| Araña and León (2013) | Corporate social responsibility decisions | III | No | A census representative sample from general population | Between | Multinomial logit | Market share prediction | Yes | - In a static setting (first month) predictions of SP experiment and real market shares were statistically similar - As time evolved, real market shares presented dynamic patterns not mirrored by the SP predictions - Differences between real market behaviour and SP data can be more attributable to the lack of learning experience or adaptation than the hypothetical nature of SP |
| Brooks and Lusk (2010) | Food products | V | No | Random sample of consumers | Within | Scaled logit | -Parameter estimates -Market share | Yes | Parameter estimates across SP and RP data could not be deemed fully congruent |
| Chang et al. (2009) | Three different private goods | II | Yes | Random sample using random digit dialling | Between | -Multinomial logit -Random parameter logit | -Parameter estimates -Market share estimates | No | - Non-hypothetical ranking method, outperformed the hypothetical CE in predicting retail sales - Non-HC's better approximation of true preferences than HC - Results suggest a high level of external validity for certain methods and models |
| Ding et al. (2005) | Chinese dinner & snacks | II | Yes | Students | Between | -Random parameter logit | -Choice probabilities | Yes | - Incentive-aligned choice conjoint outperforms hypothetical choice conjoint in out-of-sample predictions - Conventional conjoint analysis exhibits hypothetical bias |
| Liebe et al. (2019) | Ethical food consumption (tea) | II | Yes | Random sample of consumers | Between | -Scaled logit -Random parameter logit | -Parameter estimates -MWTP | Yes | Statistically significant lower WTP values for the attributes in the CE involving real payments |
| Luchini and Watson (2014) | Laboratory good with induced value | II | Yes | Students | Between | Multinomial Logit | -Proportion of Pay-off maximising choices -WTP | Yes | - CEs fail to elicit payoff maximising choices - Little evidence that increasing the salience of attribute levels, monetary incentives, or both increases the proportion of payoff maximising choices |
| Lusk and Schroeder (2004) | Food quality (Beef ribeye steak) | II | Yes | Random sample of consumers | Between | -Random parameter logit -Heteroskedastic Extreme Value -Multinomial Probit | -Parameter estimates -WTP -Choice probabilities | Mixed | - Hypothetical responses predicted higher probability of purchase - Hypothetical responses overestimated TWTP - MWTP for steak quality not different statistically across hypothetical and real payment choice |
| Mørkbak et al. (2014) | Choice of food (apple variety) | II | Yes | Random sample of consumers | Between | Random parameter (error component) logit | -Parameter estimates -WTP -Attribute processing strategies | No | Introducing a real economic incentive to a treatment sample in a CE survey had no significant effect on preferences and WTP. |
| Moser et al. (2013) | Choice of food (apple variety) | II | Yes | Random sample of shoppers | Between | -Multinomial Logit -Random parameter logit | -Parameter estimates -WTP | Yes | - Results confirm the presence of hypothetical bias and the mixed effectiveness of a cheap talk script - Estimated WTP in the hypothetical scenarios is statistically higher than the real ones. |
| Mueller et al. (2010) | Wine choice | V | No | Consumers | Within | N.A. | Market share | Mixed | - A CE without tasting was predictive for market share, while the informed tasting information was not related to sales data - A multi-media shelf simulation CE results in a better approximation for what consumers purchase in reality than the other response measures used |
| Resano-Ezcaray et al. (2010) | Food products | V | No | Random sample of consumers | Within | Nested Logit trick | -Parameter estimates -Market share | Mixed | SP model predicts general market trends but not market shares |
| Sanjuán-López and Resano-Ezcaray (2020) | Saffron | II | No* | Random sample of consumers | Within | -Random parameter logit | -Parameter estimates -MWTP | Yes | - WTP was higher in the hypothetical setting<br>- Preferences were more homogenous in the real setting<br>- WTP and magnitude of bias varied across individuals with and without knowledge about the product and its characteristics<br>- Consumers with higher WTP in hypothetical setting had also relatively higher WTP in real setting<br>- Evidence of social desirability bias observed |
| Wlömert and Eggers (2016) | Music streaming service | II | Yes | Random sample of consumers | Between | Multinomial logit | -WTP | Yes | The incentive-aligned version improved predictive accuracy |
| Wuepper et al. (2019) | Moral good (sustainable products) | III | Yes | Online consumers of a coffee shop | Between | -Multinomial logit -Random parameter logit -Generalised logit -Latent-class logit | -Parameter estimates -MWTP | Yes | The estimates from the SP experiments were confounded with the preference for appearing as "good people". |
| Yue and Tong (2009) | Food products | II | Yes | Random sample of consumers | Between | Random parameter logit | -Parameter estimates -WTP | No | When real products were used, the hypothetical bias was not high. |

*Sig. bias: Significant bias*
*HB: Hypothetical bias*
*WTP: Willingness to Pay*
*MWTP: Marginal willingness to pay*
*RP: Revealed Preference*
*\*Status-quo option was available in the choice sets*

## Appendix C: Summary of relevant studies in health economics

The table below presents a summary of studies in the discipline of health economics on hypothetical bias in choice experiments.

| Reference | Choice context | Benchmark data class | Opt-out | Sample | Within/ between sample | Estimation method | Measures of validity/bias | Sig bias? | Conclusions/comments |
|---|---|---|---|---|---|---|---|---|---|
| Kesternich et al. (2013) | Health care cover- Medicare Part D | IV | Yes | Representative sample of older Americans | Within | -Multinomial logit -Scaled logit | -Parameter estimates -WTP -Market share | No | - Hypothetical and real data produce similar estimates of WTP for insurance plan attributes<br>- Significance and signs of the estimated coefficients were always the same in the real and hypothetical data |

| | | | | | | | | | Significant scale difference |
|---|---|---|---|---|---|---|---|---|---|
| Krucien et al. (2015) | Sleep apnea treatment | IV | Yes | Patients with sleep apnea | Within | Generalised multinomial logit | -Market share -Individual choices | Yes | Results raised questions about the external validity of CE in health<br>At the sample level, the comparison showed large but not significant differences between the two methods<br>At the individual level, the comparison showed that the two methods led to significantly different patterns of choices |
| Lambooij et al. (2015) | Vaccination behaviour | IV | No | Parents of newborns | Within | Mixed logit | -Sensitivity -Specificity -Uptake | Mixed | The predictive value of the CE was satisfactory for predicting the positive choice but not for predicting the negative choice. |
| Linley and Hughes (2013) | New medicine adoption | IV | No | Appraisal committee members | Within | Random effects logit | -Sensitivity -Specificity | No | On average, appraisal committee members' stated preferences appear consistent with their actual decision-making behaviours, providing support for the external validity of our CEs |
| Mark and Swait (2004) | Physicians' prescription of alcoholism medication | IV | Yes | Sample of physicians | Within | -Multinomial logit -Scaled logit | Parameter estimates | No | Joint estimation suggests parameters from the revealed and stated preference data are equal, up to scale |
| Mohammadi et al. (2017) | Latent Tuberculosis Infection treatment | IV | Yes | Patients | Within | Mixed logit | -Individual choices (hit rate) -Sensitivity -Specificity | No | The best model correctly predicted actual treatment decisions for 83% of the participants<br>Individual-specific coefficients reflected respondents' actual choices more closely compared with the aggregate-level estimates |
| Ryan and Watson (2009) | Women's preferences for Chlamydia screening | IV | Yes | Clinic visitors | Within | Multinomial logit | Uptake rate | Yes | Significant differences between stated screening intention and actual screening uptake |
| Salampessy et al. (2015) | Diabetes intervention | IV | Yes | Diabetes patients | Within | Mixed logit | -Individual choices -Uptake -Sensitivity -Specificity | No | Both at an aggregate population level and at an individual level, high correspondence rates were found between the predicted and actual participation behaviour<br>Stated preferences derived from a CE can adequately predict actual behaviour |
| Strauss et al. (2018) | HIV testing | IV | No | Sample of truck drivers | Within | Conditional logit | Individual choices | No | Stated preference structures helped explain the actual choices made regarding the type of HIV testing |
| Telser and Zweifel (2007) | Hip protector | IV | Yes | Elderly people | Within | Probit | Uptake rate | No | Strong evidence in favour of external validity of the CE method |

Sig. bias: Significant bias
HB: Hypothetical bias
WTP: Willingness to Pay
MWTP: Marginal willingness to pay
RP: Revealed Preference

## Appendix D: Summary of relevant studies in transport economics

The table below presents a summary of studies in the discipline of transport economics on hypothetical bias in choice experiments.

| Reference | Choice context | Benchmark data class | Opt-out | Sample | Within/ between Sample | Estimation method | Measures of validity/bias | Sig bias? | Conclusions/comments |
|---|---|---|---|---|---|---|---|---|---|
| Beck et al. (2016) | Driving behaviour | III | Yes | Sydney motorists | Within | -Generalised mixed logit<br>-Joint choice and certainty estimation | -TWTP<br>-Model predictions | Yes | - Jointly estimating choice and choice certainty significantly reduces hypothetical bias<br>- Incorrect calibration of responses can produce SP results that are more biased than doing nothing |
| Brownstone et al. (2000) | Alternative-fuel vehicle choice | IV | No | Random sample of households | Within | -Multinomial logit<br>-Mixed logit<br>-Scaled logit | -Parameter estimates<br>-Model predictions<br>-Market share | Yes | Pure SP model give implausible forecasts |
| Brownstone et al. (2003) | Mode choice (solo, carpool, tolled express lane) | V | No | Morning commuters | Between | Multinomial logit | Value of time | Yes | Revealed WTP to reduce congested travel time is higher than SP results |
| Brownstone and Small (2005) | Road pricing (value of time & reliability) | V | No | Road users | Between | Mixed logit | -Value of time<br>-Value of reliability | Yes | The VoT estimated based on SP was less than half the amount estimated based on revealed behaviour |
| Fifer et al. (2014) | Driving behaviour | III | Yes | Sydney motorists | Within | -Generalised mixed logit | -MWTP<br>-TWTP<br>-Model predictions | Yes | - Hypothetical bias is a significant issue in SP surveys<br>- Mitigation techniques (cheap talk and certainty scales) have potential to compensate for the bias |
| Ghosh (2001) | Road pricing (value of time & reliability) | V | No | Morning commuters | Between | -Multinomial logit<br>-Heteroskedastic logit | Value of time | Yes | The Value of Time estimates from SP models are significantly lower than the RP estimates |
| Haghani and Sarvi (2018) | Evacuation route choice | III | No | Students | Between | -Multinomial logit<br>-Mixed logit<br>-Random regret logit | -Parameter estimates<br>-Model predictions<br>-Market share | No | - Hypothetical bias was not significant<br>- SP and RP models predicted individual choices in a fairly similarly way |
| Haghani and Sarvi (2019) | Evacuation route choice | III | No | Students | Between | -Multinomial logit<br>-Generalised mixed logit<br>-Random regret logit | -Parameter estimates<br>-System-level aggregate predictions | No | - Hypothetical bias was even less noticeable when considering aggregate system-level predictions measures such as evacuation times and exit utilisations<br>- SP and RP models produced similar system-level simulation outputs<br>- Significant scale difference between SP and RP models |

| Hultkrantz and Savsin (2017) | Time purchase | II | No | Students | Between | Probit model | Value of time saving | Yes | - Negative hypothetical bias found for allocation of time at another occasion than the present<br>- A pivot design made the magnitude of the downward bias even worse |
|---|---|---|---|---|---|---|---|---|---|
| Krčál et al. (2019) | Time purchase | II | Yes | Students | Between | Tobit model | Value of time saving | Yes | - A substantial hypothetical bias with the average SP VoT being only 70% of the corresponding RP value<br>- Participants attach a higher value to (waiting) time in the RP setting than in the SP setting<br>- Scheduling constraints are taken into account to a much lesser extent in the SP setting than in the RP setting |
| Li et al. (2018) | Mode choice | IV | No | Sample of Sydney commuters | Between | -Mixed logit | -Value of travel time saving | Mixed | - The SP VoT is significantly higher than the corresponding RP value for public transport modes<br>- For the car mode, SP delivers a similar WTP value compared to RP<br>- Pivot SP design significantly reduces the bias<br>- Presenting a full distribution of travel time in choice scenarios reduces the bias |
| Nielsen (2004) | Road pricing (value of time) | III | No | Sample of 500 driver commuters | Within | -Multinomial logit<br>-Error component logit | Value of time | Yes | Respondents tend to underestimate their behavioural changes due to road pricing in the SP experiment compared to the field experiment. |
| Peer et al. (2014) | Departure time choice | IV | No | Road users | Within | -Multinomial logit<br>-Latent-class logit | -Value of time -Value of scheduling delay | Yes | - VoT estimates derived from SP and RP data sources for the standard scheduling model were fairly close<br>- The values of early and late schedule delays differed considerably depending on whether RP or SP data was used<br>- Significant scale difference between SP and RP models |
| Small et al. (2005) | Mode choice (solo, carpool, tolled express lane) | V | No | Road users | Within | Mixed logit | -Value of time -Value of reliability | Yes | - Scale difference between SP and RP models was significant<br>- The value of time and value of reliability were substantially underestimated based on SP data when compared to RP |

*Sig. bias: Significant bias*
*HB: Hypothetical bias*
*WTP: Willingness to Pay*
*MWTP: Marginal willingness to pay*
*RP: Revealed Preference*
*VoT: Value of Time*

**Appendix E: Summary of relevant studies in psychology and decision neuroscience**

The table below presents a summary of studies in the field of psychology and decision neuroscience on hypothetical bias.

| Reference | Choice context | Benchmark data class | Within/ between sample | Experiment design factors | Sig. bias? | Conclusions/comments |
|---|---|---|---|---|---|---|
| Bickel et al. (2009) | Intertemporal choice | II | Within | - Outcome delay <br> - Amount of immediate outcome <br> - Real vs hypothetical outcome | No | - No statistical difference in the estimated measure of discounting between equal amounts of a real and hypothetical gains <br> - No significant fMRI signal change across the hypothetical and real treatments (after corrected for multiple comparisons) |
| FeldmanHall et al. (2012a) | Moral choice (between financial personal gain vs preventing physical harm to another person) | II | Within | -Real vs imaginary pain <br> -The magnitude of the pain | Yes | - Significantly more money kept in the real task <br> - Differential neural mechanisms for real and hypothetical moral decisions were identified <br> - Hypothetical moral choices mapped closely onto the imagination network, while real moral decisions recruited areas associated with social and affective processes |
| FeldmanHall et al. (2012b) | Moral choice (between financial personal gain vs preventing physical harm to another person) | II | Within | -Real vs imaginary pain <br> -The magnitude of the pain <br> -The richness of contextual information | Yes | - Real moral decisions can dramatically contradict moral choices made in hypothetical scenarios <br> - Systematic enhancement of the contextual information incrementally reduced hypothetical bias <br> - The more abstract the contextual information, the more subject's responses diverged from real behaviour. |
| Grebitus et al. (2013) | Purchase of consumer goods | II | Between | -Elicitation procedure (choice vs auction) <br> -Incentives (hypothetical vs non-hypothetical) | Yes | - Personality plays a larger role in explaining behaviour in CE than in auctions <br> - Personality could explain a significant portion of hypothetical bias <br> - Agency was the most important trait in CEs <br> - The effects of personality differ depending on whether the setting is hypothetical or real |
| Green and Lawyer (2014) | Delay and probability discounting in choice | II | Between | -Hypothetical vs real <br> -Delay <br> -Probability | Mixed | - Discounting patterns are context specific <br> - Potentially real and hypothetical discounting outcomes may be equal in monetary contexts, but not in non-monetary contexts <br> - Delay and probability discounting of potentially real cigarettes were steeper compared to hypotheticals |
| Kang et al. (2011) | Purchase of consumer goods | II | Within | Real versus hypothetical outcome | Mixed | - Activity in common areas of the orbitofrontal cortex and the ventral striatum correlated with behavioural measures of the stimulus value of the goods in real and hypothetical condition <br> - Activity in these regions was stronger in response to the stimulus value signals in the real choice condition <br> - Hypothetical and real choice differentially activate common valuation areas |
| Kang and Camerer (2013) | Aversive choice | II | Within | -Type of aversive food <br> -Real vs hypothetical | Yes | - The direction of hypothetical bias depends on whether the choice is for receiving something 'good' or avoiding something 'bad' <br> - The bias is reversed in sign for aversive choices <br> - People pay more to avoid bed outcomes when choice is real compared to when it is hypothetical. <br> - Real choice more strongly activated reward regions in the brain |
| Lawyer et al. (2011) | Delay and probability discounting in choice | II | Within | -Hypothetical vs real<br>-Delay<br>-Probability | No | - Discounting for hypothetical and potentially real outcomes yielded similar data<br>- The similarity was observed for both nicotine-dependant and non-dependant adults |
| Madden et al. (2003) | Delay and probability discounting in choice | II | Within | -Hypothetical vs real<br>-Delay<br>-Probability | No | - Discounting was the same under both hypothetical and real outcome treatments<br>- Support for the validity of using hypothetical rewards to estimate discounting rates in choice making |
| Morgenstern et al. (2014) | Choices with risky monetary outcomes | II | Within | -Sure payoff amount<br>-High-stake probabilistic payoff | Yes | - Subjects were more risk averse for real payoff choices<br>- Increased cognitive control for hypothetical decisions based on EEG signals |
| Robertson and Rasmussen (2018) | Delay and probability discounting in choice | II | Within | -Hypothetical vs real<br>-Delay<br>-Probability | No | For food-related outcomes, potentially real and hypothetical food outcomes are discounted similarly |
| Vlaev (2012) | Social dilemmas (prisoner's dilemma) | II | Between | -Real versus hypothetical<br>-Cooperation Index | Yes | - Hypothetical and real decisions cause different cognitive biases in social dilemmas<br>- Without the corrective real social interaction, people overestimate theirs and others propensity to act cooperatively |

*Sig. bias: Significant bias*
*fMRI: Functional Magnetic Resonance Imaging*
*EEG: Electroencephalogram*